**Christina Dienhart, Luis Kaufhold und Frank T. Piller**
**in collaboration with Dina Franzen-Paustenbach and Michael Schmitt**

# Urban Metaverse: The Smart City in the Industrial Metaverse

## Opportunities of the Metaverse for Real-Time, Interactive, and Inclusive Infrastructure Applications in Urban Areas

A trend study by the Institute for Technology and Innovation Management, RWTH Aachen University, in cooperation with regio iT GmbH

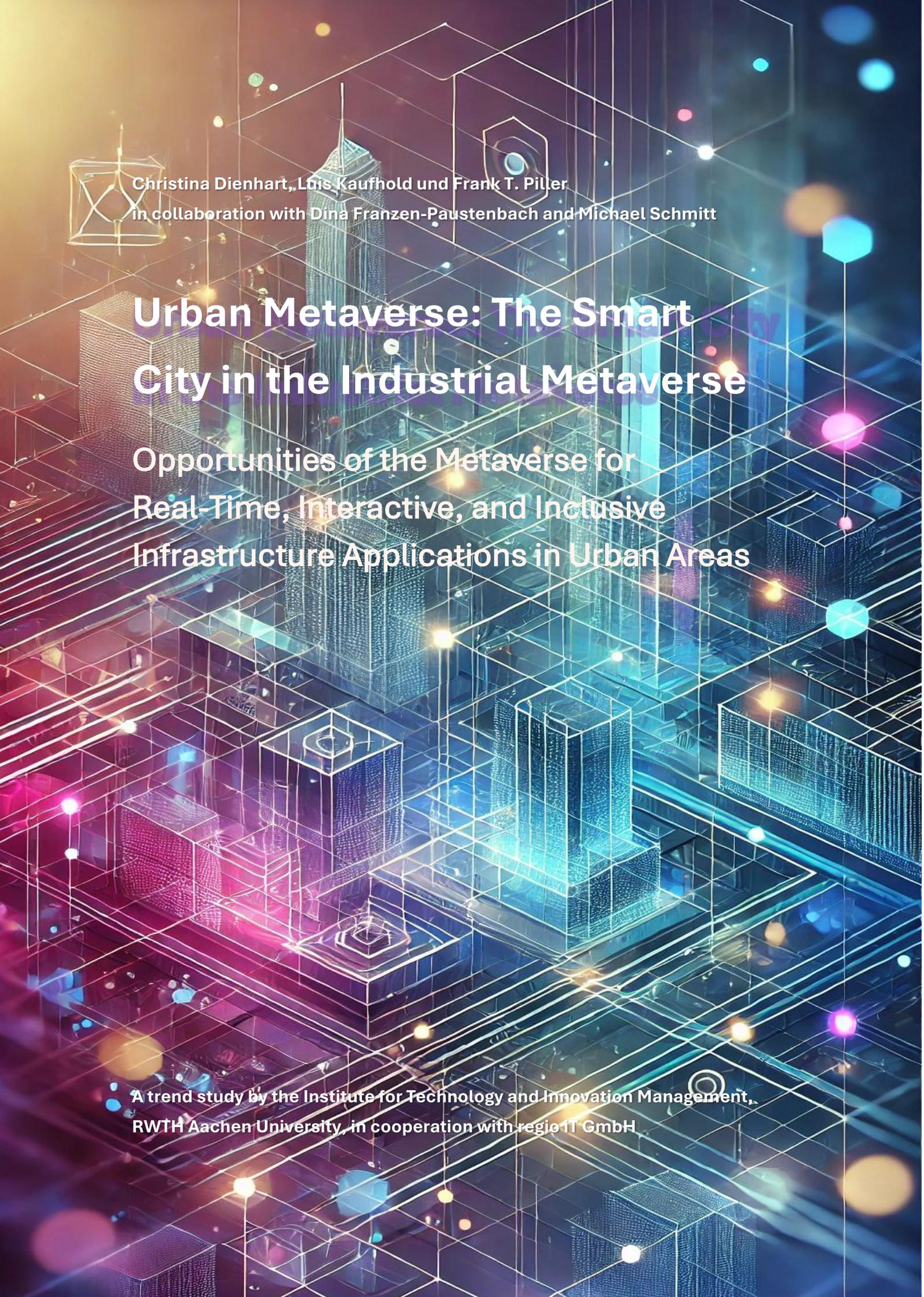

# Urban Metaverse: The Smart City in the Industrial Metaverse

Opportunities of the Metaverse for Real-Time, Interactive, and Inclusive Infrastructure Applications in Urban Areas

A trend study by the Institute for Technology and Innovation Management, RWTH Aachen University, in cooperation with regio iT GmbH

# Welcome to the Urban Metaverse

In the context of the smart city, the urban metaverse represents an immersive 3D environment that seamlessly integrates the physical city with its digital data and systems. By merging physical and digital realities, it opens up new possibilities for urban planning, development, and everyday life.

Our study explores these opportunities, examines the key technologies driving the urban metaverse, and critically analyzes the socio-economic challenges of its implementation. The focus is on how the urban metaverse can optimize the planning and operation of urban infrastructures, enhance inclusion and citizen participation, and strengthen the innovative capacity of cities and municipalities.

Four of the study's recommendations for the implementation of metaverse applications in an urban context are highlighted.

**1. User-Centered Design:** Applications in the urban metaverse must be designed to address specific, currently unmet needs and requirements of a city's key stakeholders—including citizens, municipal administrations, and the commercial sector—by offering a clear and innovative value proposition. Successful urban metaverse development requires a holistic approach, ensuring that its applications create tangible benefits across all user groups while enhancing the overall functionality and sustainability of the smart city.

**2. Ubiquitous Accessibility:** The urban metaverse should be universally accessible, ensuring low-threshold entry and barrier-free participation for all citizens, regardless of their technical proficiency, financial resources, or physical abilities.

**3. Proactive Design of the Legal Framework:** The urban metaverse generates vast amounts of user data, raising not only concerns about data protection and security but also new legal questions—such as the legal and business capacity of avatars or liability for prescriptions from a digital twin. Cities must take a proactive role in shaping the legal framework, viewing legal considerations not as obstacles but as catalysts for innovation.

**4. Development of Sustainable Business Models:** The implementation and operation of an urban metaverse require significant investment, making the early design of viable business models essential. To support this, we provide dedicated business model templates, including approaches for data monetization, platform ecosystems, and freemium or subscription-based virtual services.

Our study aims to serve as both a source of inspiration and a practical guide for decision-makers in cities and municipalities, urban planners, IT experts, local business leaders, and anyone interested in the future of urban spaces. It provides a comprehensive understanding of the opportunities and challenges of the urban metaverse as an evolution of the smart city, helping to set the course for sustainable and innovative urban development.

The study is the result of a joint research project between the Institute for Technology and Innovation Management at RWTH Aachen University and regio iT. First and foremost, we would like to thank **Dieter Rehfeld**, a former managing director of regio iT, for his inspiration and motivation for this study. We also thank **Dina Franzen-Paustenbach** and **Michael Schmitt** from regio iT's innovation management group for their regular input and constructive feedback. At PwC Germany, we would like to thank **Jil Aline Villinger**, Senior Associate in Digital Transformation & Energy, and **Philipp Schmidt**, Partner in E2E Digital Transformation, for their valuable input during the initial conceptualization phase. We thank **Heiko von der Gracht**, Professor at the Chair of Futures Studies at Steinbeis Hochschule, **André Henke**, Head of the Digital Administration Program Lower Saxony, **Harmen van Sprang**, (Co-)Founder of the Sharing Cities Alliance Foundation, for the fruitful discussions, and **Uwe Rechkemmer**, Senior Sales Specialist Simulation & Visualization for the Omniverse platform at NVIDIA, as well as **all the other experts** listed in the text who helped us with information and inspiration for this study.

Aachen, April 2025

*Christina Dienhart, Luis Alexander Kaufhold und Frank Piller*

# Content

# 1. Introduction to the Metaverse: A New World Emerges

The metaverse is a broad term encompassing the next stage of internet evolution. Unlike today's internet, which relies primarily on websites and apps, the metaverse will take the form of an immersive, three-dimensional space where the boundaries between the virtual and physical worlds blur. Through augmented reality (AR) or virtual reality (VR) glasses, users will be able to immerse themselves in digitally enhanced or fully virtual environments, exploring these spaces and interacting with others or their digital representations in for of avatars.

## The Metaverse as the Third Evolutionary Stage of the Internet

The metaverse is not merely the next iteration of social networks, as is often claimed. Instead, it represents an open, interoperable, and social ecosystem of various technologies, enabling multimodal participation across different devices. This marks the **third evolutionary stage of the internet**.

Rather than focusing on the nuances of different definitions (for this, see Ritterbusch & Teichmann, 2023), we conceptualize the metaverse—particularly in its original domain of private consumers—as a new computer-based environment composed of virtual "worlds," where people interact and communicate in real time through avatars. For users, this fosters a sense of immersion in a persistent digital space that extends beyond singular interactions (Hennig-Thurau et al., 2023). The box below provides a more detailed explanation.

What we are experiencing today is comparable to the emergence of the mobile internet in the early 2000s. Back then, the shift was not merely a matter of transferring the existing internet onto mobile devices; rather, it gave rise to entirely new devices, services, and user experiences that were fundamentally different from the traditional "stationary" internet accessed via home or office computers—many of which would have been unimaginable before. The mobile internet became the foundation for the business models of some of the world's largest technology companies, including Apple, Google, Uber, DHL, and Amazon (Gao et al., 2024; Mystakidis, 2022; Dolata & Schwabe, 2023).

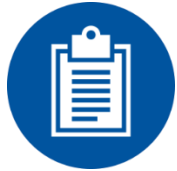

**Definition (Consumer) Metaverse**

*The (consumer) metaverse is a new computer-based environment consisting of virtual "worlds" in which people interact and communicate with each other in real time via avatars. For users, this creates a feeling of immersion in this world, which also continues over time.*

*The metaverse consists of platforms and devices that work together seamlessly and allow thousands of people and machine agents to interact with each other simultaneously. The metaverse is thus an inherently social place and provides space for a wide range of shared human activities, from entertainment to collaboration at work. Access to this is provided by specific hardware (e.g. AR/VR headsets) as well as dedicated operating systems and special apps.*

Just as the mobile internet has become the norm today, we see the metaverse as the next evolutionary stage of the internet — introducing new devices and entirely new applications while remaining grounded in the same fundamental principles and technological foundations. In terms of user interfaces for interacting with the metaverse, we are still at an early stage, comparable to the first Nokia cell phones with mobile internet. The equivalent of modern smartphones — such as the iPhone — is still missing, despite advancements like Apple's Vision Pro, Meta's Quest 3, or Ray-Ban's Headliner. It would therefore be a mistake to judge the future of metaverse interactions based on today's bulky VR headsets. Significant advancements are still to come, and only when these interaction technologies mature will the metaverse achieve widespread adoption.

Although the metaverse is still in its early stages of development, its potential applications are vast and diverse. Given the broad and sometimes ambiguous definition of the metaverse, there is ongoing debate about where it truly begins for certain applications. In the realm of private consumers—often referred to as the **consumer metaverse**—key **use cases** include the following (Hadi et al., 2024; Hennig-Thurau et al., 2023):

- **Social Interaction and Gaming in Virtual Worlds:** The metaverse was originally envisioned as a virtual space where people from around the world could connect— whether to share a virtual coffee in a meticulously recreated Parisian brasserie or to engage in other immersive social experiences from the comfort of their homes. The goal is to foster a sense of connectedness and presence, even across vast distances (Hennig-Thurau et al., 2023; Ritterbusch & Teichmann, 2023).

    Platforms like Meta Horizon Home already enable users to meet friends, watch TV together, and play games in shared virtual environments. Among the most developed use cases of the metaverse today is gaming, where users immerse themselves in expansive digital worlds. Platforms such as Roblox, Fortnite, and Minecraft already offer metaverse-like experiences, blending gaming with social

interaction, virtual economies, and customizable avatars. Many online games are incorporating new features that merge gaming with reality, making them even more immersive. Additionally, VR technology has found its widest adoption in the gaming sector, enabling deeply engaging experiences that push the boundaries of traditional gameplay and social interaction in virtual environments.

- **Consumption and Digital Ownership:** Some experts believe that a complete digital economy will be established in the metaverse, in which virtual goods, services (and perhaps even real estate) can be traded on the basis of NFTs, i.e. non-fungible tokens. Tokenization makes it possible for these virtual assets, just like physical objects, not to be copied and to have a unique originator (Allam, Sharifi, Bibri, Jones & Krogstie, 2022; Hözle et al., 2023).

  Their value increases with increasing interoperability, which allows them to be transferred to different metaverse platforms (Hvitved et al., 2023; Marabelli, 2023). Assets do not necessarily have to have a pedant in the physical world. There is already a large market for gaming skins that change the appearance of avatars and their virtual objects. We believe that virtual assets that have no physical equivalent and a high self-actualization value have economic value (Hasan et al., 2024; Yoo et al., 2023).

- **Health, Wellness and Telemedicine:** The metaverse also holds potential for telemedicine, where patients can interact with doctors in virtual clinics or consultation rooms or take part in virtual health and fitness programs (Shardeo et al., 2024; Wang et al., 2022). For example, doctors can use three-dimensional images or simulations to explain complex medical issues in an understandable way. The idea of offering personalized, interactive physiotherapy or a rehabilitation program after an operation is also exciting.

  Patients can complete exercises in a virtual environment that motivates and encourages them, while their progress is monitored in real time using VR systems with motion sensors. This is also conceivable for therapies against trauma, stress or mental illness. Here, patients can complete meditation or mindfulness exercises in specially designed, calming environments to support the mental healing process. Therapists can offer psychological support in virtual sessions.

## The Metaverse in an Industrial and BtoB Context

In the private sector, enthusiasm for the metaverse has somewhat cooled following the hype of 2021 and 2022. This decline is largely due to insufficient mobile network infrastructure and the immaturity of the required hardware needed to achieve the seamless integration of digital and physical worlds.

However, the **industrial metaverse** tells a different story. Since mid-2023, this concept has gained significant traction in professional and industrial circles, driven by companies such as Siemens, NVIDIA, and BMW. While the term "industrial metaverse" remains loosely defined, it broadly refers to a virtual universe that connects physical and digital environments, enabling real-time interaction between humans and machines, seamless data exchange, and the provision or use of services in an immersive, interoperable setting (Enders et al., 2024; Zhang et al., 2024; Hasan et al., 2024).

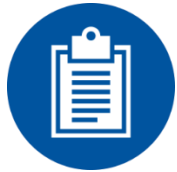

**Definition Industrial Metaverse**

*The industrial metaverse can be defined as a virtual 3D environment that connects digital twins of real industrial objects, processes and systems.*

*This digital representation enables real-time interaction and analysis to improve efficiency, productivity and sustainability in industry.*

We see the industrial metaverse as a **platform that integrates various stand-alone industrial systems** into a unified, interoperable environment (Springer et al., 2025). Today, when a plant is planned, it is first modeled and simulated using specialized CAD programs. The construction phase then relies on separate project management systems, while the operational phase involves yet another set of tools for training, maintenance, and continuous improvements.

A key vision for the industrial metaverse is to consolidate these fragmented applications onto a single, standardized platform, where digital models can be seamlessly combined and accessed. This would enhance efficiency, collaboration, and accessibility across the entire industrial lifecycle. A crucial enabler for this is the Universal Scene Description (USD) standard, which we will explore in more detail in Chapter 3. The industrial metaverse offers a wide range of applications, with the following key **use cases** being particularly relevant:

- **Remote Work:** Virtual meetings and conferences take place with avatars in the metaverse from anywhere. Unlike online meetings in Microsoft Teams or Zoom, which are characterized by "screen fatigue" (Bailenson, 2021; Kuru, 2023), participants can really interact with each other here - and the VR headset goes from being a limitation to a "feature", as it forces participants to focus on the meeting and not multi-task on a second screen (Marabelli, 2023). The metaverse thus becomes a platform for collaboration and the exchange of ideas between colleagues, teams, departments and even companies. The most competent employee can also be connected at any time for customer concerns, which significantly improves customer support (Hölzle et al., 2023; Marko et al., 2023). Similarly, excursions are already commonplace today: at the beginning of 2023, one of the authors (Frank Piller) was able to visit the first works council office of IG Metall in the metaverse - which, incidentally, takes a very constructive approach to the topic and proactively engages with the work in the metaverse (IG Metall, 2023).

- **Recruitment:** At the RWTH Aachen University, for example, we are already seeing more and more companies offering their recruitment events and job fairs in the metaverse (or at least in VR environments) in the battle for the best graduates - not only because this offers completely new options for presenting companies, but also to get in touch with students in a low-threshold but very interactive and personal way (and at significantly lower costs than at an on-site event).

- **Training and Education:** Immersive learning experiences in the metaverse, powered by AR and VR technologies, enable learners to engage and practice in realistic environments. Interactive 3D models and simulations enhance understanding and facilitate knowledge exchange, particularly for complex objects such as machines and industrial processes. Research indicates that immersive learning significantly improves learning outcomes, with a key advantage being the high level of interaction between learners. Additionally, the flexibility of time- and location-independent learning allows for individualized training tailored to employees' specific needs. At the same time, companies can achieve cost savings compared to traditional in-person training programs.

- **Manufacturing and Production Planning:** In the manufacturing industry, digital twins of machines and systems are connected in virtual factories in the metaverse and can be used for advanced modeling and simulations. Before planning is implemented in physical reality, many more iterative trial-and-error processes are possible by running through various scenarios by the players involved. A new factory can already be visited during the planning phase in order to collaboratively validate setups or identify optimization requirements. This allows specialist staff to identify and correct errors in the planning before the actual construction begins, while customers get a better impression of the end result during a virtual tour. The factory can then be operated virtually with all employees in the metaverse for a few months and improved there before it goes "live" (Hölzle et al., 2023) - but then immediately with a higher start-up capacity. From this point onwards, the system status can then be monitored and predicted. Overall, this can improve the efficiency and precision of production processes. BMW's activities in the Omniverse (NVIDIA's industrial metaverse platform) are moving in precisely this direction.

- **Supply Chain Management:** In the metaverse, the concept of digital modeling, simulation, and real-time optimization can be extended beyond individual companies to entire supply chains. By creating digital twins of supply networks, businesses can visualize and monitor operations in real time, enabling more accurate forecasting and control of supply chain events. This digital integration allows companies to anticipate disruptions, optimize resource allocation, and enhance collaboration across supply chain partners. As a result, businesses can reduce waste, lower costs, and improve supply chain resilience, ensuring greater adaptability in the face of unexpected disruptions or crises.

- **Sales:** With virtually limitless opportunities for brand promotion and product presentation in virtual showrooms, the metaverse serves as an ideal platform for marketing and commerce (Arunov & Scholz, 2023). As a virtual marketplace, it has the potential to seamlessly connect manufacturers, customers, partners, and suppliers, fostering new business models and interactive shopping experiences (Hölzle et al., 2023; Marko et al., 2023).

The use cases demonstrate that the metaverse has the potential to fundamentally transform business operations by enhancing efficiency, enabling seamless remote collaboration, and improving user experiences. By integrating virtual environments into workflows, companies can streamline processes, reduce operational costs, and foster more engaging interactions, ultimately driving innovation and productivity.

While the literature on the industrial metaverse (with almost 600 publications on Google Scholar as of the end of 2024) and the metaverse in general (over 36,000 publications) is already quite extensive, one domain remains relatively underexplored—despite its significant potential: the intersection of smart cities and the metaverse.

We believe that a "Smart City metaverse" – or, as we call it, the **urban metaverse** (Hudson-Smith & Batty, 2024; Kuru, 2023) – will evolve in a manner similar to the industrial metaverse, albeit with a time lag. Just as the concept of "Industry 4.0" laid the groundwork for the Smart City, we anticipate that the urban metaverse will follow the industrial metaverse, integrating digital and physical urban environments to transform city life. After a brief introduction and an overview of the smart city concept in the next section, Chapter 3 will provide a more detailed analysis of the urban metaverse.

## On the Way to the Urban Metaverse in the Smart City

The **urban metaverse,** within the smart city context, builds upon the ideas and visions of both the **consumer metaverse** and the **industrial metaverse**. It represents an immersive 3D environment that seamlessly connects a city's physical spaces, citizens, and digital systems in real time. This fusion of physical and digital realities unlocks new possibilities for urban design, interaction, and services. Access to the urban metaverse can be facilitated through VR or AR glasses, but also via traditional devices such as PCs, laptops, or smartphones, ensuring broad accessibility.

This integration enables a wide range of complementary use cases, generating positive spill-over effects across different urban functions. For instance, a city can enhance its digital presence by positioning itself as a brand and attractive destination within the metaverse while simultaneously fostering more inclusive and dynamic interactions. The following examples of **urban metaverse use cases** illustrate these opportunities in greater detail.

- **Shopping:** Due to the migration of customers to internet platforms such as Amazon or eBay, city centers have been suffering for years from the desolation of once flourishing shopping areas. Initiatives to merge the offline experience on site with digital offerings from large platforms such as Google Maps or local platforms such as Smart Shopping **Aachen** serve as a counter-movement. Here, local retailers and restaurants can advertise their products, arrange appointments and provide a glimpse into their business through images and 360 degree shots. Such initiatives are a good first step, especially for small, owner-managed stores, but they are not enough. By expanding the shopping experience through AR and replication in VR, stores are given the opportunity to present their existing physical offerings to a broader customer base. When shopping in the physical world, clothes can now also be tried on in the mirror or on an AR monitor with AR glasses before purchase. Companies such as ZERO10 already offer this solution today. If the weather is not conducive to shopping, it is also possible to fit clothes on personalized avatars. Furniture and decorations can also be placed in the personal home (also in the metaverse) on a test basis. A constantly available and, thanks to AI, competent shopping advisor is on hand.

- **Tourism:** The shopping experience in the urban metaverse becomes even more immersive when customers can simultaneously explore the city's cultural and historical landmarks virtually. Parks, museums, and historic buildings can be digitally recreated, allowing visitors to experience them from anywhere. For cities struggling with overtourism, the metaverse provides a way to preserve cultural heritage while still making it accessible to a global audience (Allam et al., 2022). Through virtual models, art monuments and otherwise restricted historical sites become digitally accessible, ensuring a broader reach (Deutscher Städtetag, 2024). The metaverse also enables historical reconstructions, offering visitors the chance to take a journey back in time — for instance, experiencing **Aachen** during the reign of Charlemagne or exploring a medieval version of today's urban landscape. For on-site visits, augmented reality (AR) applications can further enrich exploration. Restoration work on buildings such as **Cologne** Cathedral could be visually removed, allowing visitors to see the monument as it was originally designed. Customizable AR filters could also alter the urban aesthetic, either removing graffiti for those who prefer a cleaner look or transforming pedestrian zones into dynamic street art displays based on user preferences.

- **Education:** A metaverse for education creates opportunities for citizens to learn from university staff, international subject matter experts, and even from each other in a collaborative environment (Arunov & Scholz, 2023). This peer-to-peer classroom setting is particularly valuable for individuals with limited time or restricted access to traditional education, such as single parents or residents of rural areas (Marabelli, 2023). Similar to adult education courses, the metaverse can accommodate a wide range of topics across multiple virtual learning spaces. By incorporating 3D learning content, complex concepts—such as human anatomy or knot tying—can be made more intuitive and engaging, enhancing knowledge acquisition and retention (smartphone apps already offer similar functionalities for these examples). A virtual toolbox further enables users to interact playfully with 3D content, for instance, by tying knots directly in three dimensions. The range of educational formats is vast, from pottery courses to escape games, with the added advantage of VR and AR integration for practical, hands-on learning. Embedding knowledge into real-world application contexts makes education more immersive and relevant, allowing problem-solving formats to address both individual learning needs and city-wide challenges in a highly interactive way (Hölzle et al., 2023).

- **Inclusion and Citizen Participation:** The urban metaverse also presents new opportunities to enhance inclusion and citizen participation by expanding access to political processes and simplifying public interaction (Datta, 2022a; Hvitved et al., 2023). Through barrier-free and flexible participation formats, it enables greater involvement, particularly for people with physical disabilities and socially disadvantaged groups. Key features such as multilingual capabilities and culturally diverse virtual spaces promote inclusion across different population segments. Public meetings in the metaverse allow citizens to attend city council sessions and participate in local governance from anywhere, removing location-based barriers. Virtual consultation hours with politicians, along with gamified participation formats, foster dialogue and transparency in decision-making processes. The metaverse also supports grassroots initiatives, making it easier to organize civic movements and conduct virtual voting. Additionally, immersive urban planning tools and simulations help citizens better understand proposed projects, enabling them to provide more informed feedback. Overall, by offering innovative digital participation formats, the urban metaverse strengthens participatory democracy, making political engagement more inclusive, transparent, and accessible.

These exemplary use cases illustrate that the urban metaverse is far more than just a VR showcase in a city museum (Geraghty, Lee, Glickman & Rainwater, 2022; Hudson-Smith & Batty, 2024; Kuru, 2023). It represents a fundamental transformation in how cities operate, engage citizens, and integrate digital innovation into urban life. **The goal of our study** is to analyze both the opportunities and potential of the urban metaverse, while also addressing the challenges and implementation hurdles that cities must overcome to successfully adopt this technology:

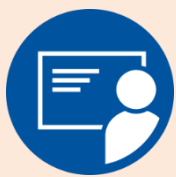

*The questions leading our inquiry for this report were: What is the current state of implementation of the urban metaverse? What are the key technological foundations of the urban metaverse, and which design factors are critical for its success? What potential does the urban metaverse offer for cities and municipalities, as well as for municipal service providers and IT companies?*

*Luis Kaufhold, Christina Dienhart and Frank Piller*

In a "deep dive," we will closely examine a typical use case of the urban metaverse within the Smart City context: the planning, structural implementation, and operation of urban buildings and infrastructure. Before diving into this specific application, we will first provide an overview of the Smart City debate (Chapter 2) and then contextualize its connection to the metaverse by exploring how the concept of the urban metaverse is shaping the future of digitally integrated cities (Chapter 3).

# 2. Smart City – the Intelligent and Participative City of the Future?

Cities are tasked with successfully managing urban transformation while simultaneously contributing to environmental and climate protection. Many see the smart city concept as a viable solution—an intelligent, participatory, and sustainable urban model. A smart city proactively addresses ecological, economic, and social challenges by ensuring that urban resources are allocated efficiently, fairly, and sustainably, all while responding to emerging needs in a timely manner.

Broadly defined, a smart city is an urban environment that leverages advanced digital technologies and innovative solutions across society, business, and government to enhance the quality of life in public spaces, improve the efficiency of urban services, and foster sustainability (Energie Baden-Württemberg [EnBW], 2024; PwC, 2024). It integrates the principles of sustainable urban development with the opportunities of digitalization, balancing the economic, ecological, and social dimensions of sustainability.

Many cities and municipalities are actively pursuing digital transformation, working to modernize key functions and processes of urban life. By embracing digital solutions, they aim to unlock new efficiencies, improve citizen engagement, and maximize the benefits of the digital world for both local governments and residents.

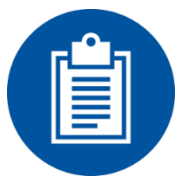

**Definition Smart City**

*A smart city leverages advanced digital technologies and innovative solutions to enhance citizens' quality of life, improve the efficiency of urban services, and drive sustainability across the economic, environmental, and social spheres. By integrating urban development strategies with the opportunities of digitalization, it creates a more responsive, adaptive, and resource-efficient urban environment. Crucially, a smart city also fosters inclusive participation, ensuring that all citizens—including those without technical expertise—can actively contribute to shaping the city's future through collaborative and accessible decision-making processes.*

This also includes citizens who are less tech-savvy or lack barrier-free access to digital technologies (German Advisory Council on Global Change [WBGU], 2019). The goal is to ensure active social participation and co-creation in urban transformation processes, enabling inclusive engagement regardless of an individual's digital proficiency or access to technology.

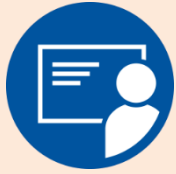

*"A smart city takes up the Industry 4.0 concept from the business world."*

*André Henke, Program Manager Digital Administration Lower Saxony (DVN), Lower Saxony Ministry of the Interior and Sport*

From a technological perspective, the smart city is often regarded as the urban counterpart to Industry 4.0 in the manufacturing sector. However, technology should always be seen as a means to an end, rather than an end in itself.

With the advancement of Internet of Things (IoT) technologies, cities can collect data from all urban domains, make it publicly accessible through open data portals, and leverage data analytics to extract insights. This, in turn, enables the development of value-added services that enhance urban life, governance, and sustainability for society as a whole.

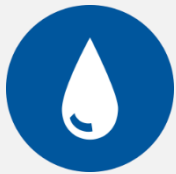

Example:
Intelligent Wastewater Network
**The Project RIWWER**

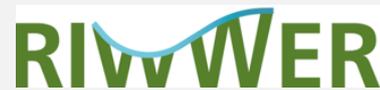

The RIWWER project aims to digitize urban wastewater networks using IoT sensors to measure water levels and flow rates. By networking with existing actuators such as pumps and weirs, a comprehensive AI-based monitoring and control system is to be developed. This will not only ensure greater transparency in the network, but will also intelligently distribute the water volumes in the network. Backwater and flooding can thus be avoided and environmentally harmful wastewater discharges reduced. Overall, it is easier to react to critical heavy rainfall events. The system therefore scores points in all areas of sustainability: social, ecological and economic.

Source: www.riwwer.org

## Areas of a Smart City

Cities and regions will only become truly "smart" when technological advancements are holistically integrated across all sectors (PwC, 2024). This requires the networking of municipal infrastructures such as energy, buildings, transportation, water, and wastewater systems, ensuring seamless interaction and optimization.

Additionally, administrative processes—both internally within city governments and externally between the administration and citizens—are improved through digital transformation. This not only enables resource conservation and enhances citizen well-being (Federal Office for Information Security [BSI], 2022), but also strengthens the city's economic competitiveness and fosters a resilient local economy.

By aligning with the sustainability policies of federal, state, and local governments (Federal Government, 2024), smart cities can become key drivers of sustainable urban development. A smart city can typically be divided into eight key areas, as outlined in Table 1 below (Federal Ministry of Housing, Urban Development and Building [BMWSB], 2024; Messe Berlin & Bitkom, 2024).

| Table 1: Target and Impact Areas of the Smart City | | |
|---|---|---|
| **Area** | **Background and Targets** | **"Smart" Technology Examples** |
| Smart Energy & Environment | ▪ Use energy efficiently and optimize energy consumption<br>▪ Integrate renewable energy sources<br>▪ Promote sustainability, e.g. reduce the city's ecological footprint | ▪ Renewable energy technology<br>▪ Smart-grid-systems<br>▪ Technology for energy efficiency in buildings<br>▪ Intelligent environmental monitoring systems |
| Smart Living | ▪ Improve quality of life by increasing comfort, safety and efficiency for citizens | ▪ Connected IoT sensors and devices for monitoring indoor climate and security<br>▪ Intelligent building automation systems<br>▪ Platforms for the management of services |
| Smart Infrastructure | ▪ Modernize and optimize urban infrastructure to ensure reliable, efficient and flexible provision of services and resources<br>▪ Meeting the needs of the growing urban population | ▪ Connected IoT sensors and actuators, big data analysis and cloud computing for controlling communication networks, supply and disposal networks and other basic services |
| Smart Health Care | ▪ Improve health care<br>▪ Facilitate access to healthcare services<br>▪ Making medical services more efficient and of higher quality | ▪ Remote monitoring systems<br>▪ Telemedicine<br>▪ Electronic patient file<br>▪ Health apps |
| Smart Governance | ▪ Make public administration processes more transparent and citizen-oriented<br>▪ Promote citizen participation for improved decision-making and innovation | ▪ E-government platform<br>▪ Digital formats for citizen participation<br>▪ Open data portals |

| Area | Background and Targets | "Smart" Technology Examples |
|---|---|---|
| Smart Transport & Mobility | ▪ Increase the efficiency of the transportation system by improving traffic flow and reducing congestion<br>▪ Facilitate the mobility of citizens<br>▪ Promote sustainable transportation<br>▪ Reduce environmental pollution | ▪ Intelligent transportation systems that monitor and control traffic in real time<br>▪ Car-sharing services<br>▪ Electromobility |
| Smart People | ▪ Promote citizens' education and digital skills to enable active participation in urban life<br>▪ Driving urban development through innovation | ▪ Educational programs<br>▪ Technology centers<br>▪ Digital inclusion initiatives |
| Smart Economy | ▪ Creation of jobs<br>▪ Increase the prosperity of citizens<br>▪ Promote entrepreneurship, innovation and economic growth<br>▪ Strengthening the city's competitiveness | ▪ Creation of a supportive business ecosystem<br>▪ Promotion of start-ups and cross-sector knowledge transfer<br>▪ Open innovation platforms |

**Table 1: Target and Impact Areas of the Smart City**

## Challenges and Current Status

While the vision of a digitally "intelligent" city—one that provides value-added services—is frequently discussed and supported by well-developed conceptual ideas, widespread practical implementation remains limited. This gap between vision and execution is also reflected in the Smart City Index, which ranks major German cities based on their level of digitalization (see Figure 1).

Experts from Bitkom Research collect, review, and qualify data points across five key categories, assessing factors such as online citizen services, shared mobility solutions, intelligent traffic light systems, and broadband infrastructure. Since 2023, education has also been included as an evaluation criterion. In the 2024 ranking, **Munich** retains its top position, followed by **Hamburg** in second place and **Cologne** in third (Bitkom, 2024). **Aachen**, which was previously among the top five, drops to 11th place.

Notably, the metaverse is not explicitly mentioned in the latest edition of the Smart City Index. However, the study does highlight significant advancements in IT & communication across major German cities in 2024, compared to 2023—laying essential groundwork for future metaverse applications.

Top 20 Cities in "Smart City Index" 2024

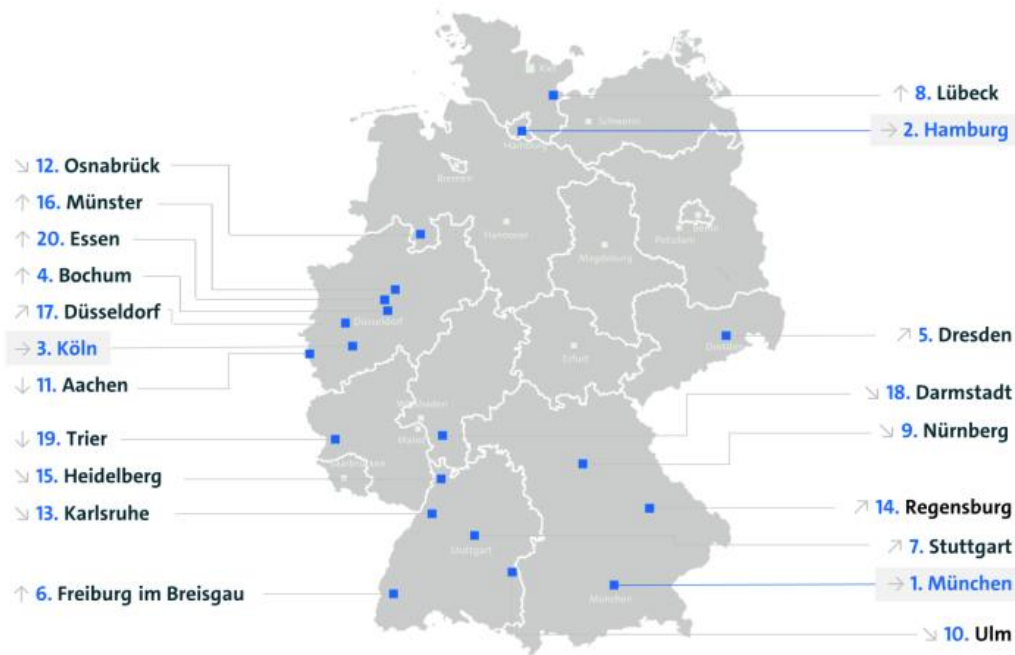

**Figure 1: bitkom's "Smart City Index" of major German cities according to their degree of digitalization** (Source: bitkom, 2024)

One of the key drivers behind this progress is the growing importance of data. According to the report:

- 79% of cities now offer open data portals, making municipal data more accessible.
- Internet and mobile network expansion is advancing, with 97% of households covered by 5G and 83% connected to broadband (Bitkom, 2024).

These developments signal important steps toward a more connected, data-driven urban environment, potentially enabling future integration of metaverse technologies into smart city infrastructures.

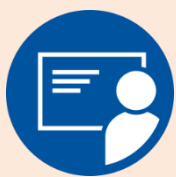

*"The smart city of tomorrow will not be located exclusively in the metaverse. Many citizen services will be classically digitized and not necessarily take place in the metaverse. I see the urban metaverse as another experience or interaction channel."*

Prof. Heiko von der Gracht, Chair of Futures Studies, Steinbeis Hochschule

The widespread implementation of the smart city concept still faces numerous barriers, ranging from financial constraints and legal uncertainties to unclear business models and immature technologies. One of the most significant challenges is establishing a sustainable and self-sustaining ecosystem that effectively connects key stakeholders, including government, industry, academia, and civil society.

A fully integrated smart city can only succeed if all stakeholder groups actively collaborate in shaping the city of the future. A common obstacle is silo thinking within different urban sectors, where innovative ideas are developed, but their implementation fails due to insufficient societal support and lack of financial resources. Without cross-sector collaboration, many promising initiatives remain isolated pilot projects rather than becoming integrated urban solutions. Hence, in the short and medium term, the trans-formation of existing urban areas into truly comprehensive smart cities remains a significant challenge (PwC, 2024). But we believe that developing a common vision of an urban metaverse could become a true game changer in this regard, as also the pioneering urban metaverse initiatives introduced in the next chapter demonstrate,

# 3.   The Metaverse in the Smart City Context

The metaverse in the smart city context, the **urban metaverse**, extends the physical city by integrating a virtual 3D environment that connects its inhabitants, digital data, and systems. This environment can be explored multimodally, with AR and VR glasses enabling users to immerse themselves in a digital representation of the city.

The urban metaverse unlocks new possibilities for urban design and service utilization, spanning areas such as entertainment, tourism, education, retail, and citizen services. Additionally, it enhances the planning, operation, and maintenance of urban infrastructures by fostering collaboration among commercial entities, city officials, and other stakeholders — creating a more efficient, data-driven, and participatory urban ecosystem.

## Aim and Essence of the Urban Metaverse

The urban metaverse, as an evolution of the smart city concept, aims to enhance efficiency, optimize resource management, foster citizen participation, and create an innovative, interactive urban environment that improves overall quality of life (Bamberger et al., 2025; Hudson-Smith & Batty, 2024; Kuru, 2023). The following definition serves as the foundation for our further exploration of this topic.

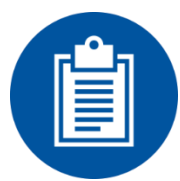

**Definition Urban Metaverse**

*The Urban Metaverse (Smart City Metaverse) describes an immersive 3D environment that connects the physical world of the city and the citizens living in it with digital twins of infrastructures, products, people and processes virtually in real time.*

*It enables a fusion of physical and digital reality that opens up new possibilities for the design and use of the city and the collaboration of all stakeholders. VR or AR glasses can be used as interfaces, as well as traditional end devices such as a PC, laptop or smartphone.*

The essence of the urban metaverse is based on the combination of the visions of the consumer metaverse and the industrial metaverse. From an overarching perspective, **the urban metaverse has the following key characteristics**:

- **Digital Twins**: Virtual copies of physical objects, processes and systems in the city that are constantly updated and linked together in a virtual environment.
- **Real time and Temporal Updating**: A fundamental aspect of the urban metaverse for users is the temporal updating of the virtual world with the physical world (persistence). Depending on the use case, this happens in real time or at regular intervals.
- **Immersion:** VR and AR technologies enable an immersive experience that allows users to immerse themselves in the virtual environment.
- **Interactivity:** Citizens, administration and companies can interact in the virtual environment, carry out simulations with objects and systems and make data-driven decisions.

It is important to highlight that **our understanding of the urban metaverse differs from the concept of the Cityverse** (Martinez, 2023), a term introduced by the European Commission to describe the interconnection of multiple urban metaverses into a larger digital ecosystem. The vision behind the Cityverse is to create an interoperable European digital twin infrastructure, linking individual urban metaverses across cities into a **unified framework**. This Cityverse is envisioned as a secure, European digital infrastructure, aligned with initiatives such as Gaia-X. This infrastructure is built on decentralized data spaces, facilitating cross-organizational and cross-border networking (Coenen et al., 2021).

In our understanding, the Cityverse represents the **interconnection of multiple urban metaverses** (or advanced smart city infrastructures) into a larger, interoperable digital ecosystem. But currently, these concepts remain vague, with no universally accepted definitions or clear distinctions between them. It is still too early to determine which term will ultimately dominate the discourse. Therefore, it is essential to always specify the context when discussing the metaverse in an urban setting, particularly in relation to specific use cases and applications. And while terminology matters, content is king, as the following sections will show.

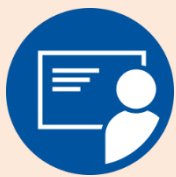

*"The EU started dealing with virtual worlds much earlier than most national governments. Strategies are being adopted on how citizens can be involved in a future EU metaverse."*

André Henke, Program Director Digital Administration Lower Saxony (DVN), Lower Saxony Ministry of the Interior and Sport

# The Pioneers: Urban Metaverse Initiatives

Ultimately, the metaverse in the Smart City will not be realized through vision papers alone, but through concrete implementations that provide tangible benefits for all stakeholders. In recent years, several cities have gained attention with high-profile urban metaverse initiatives, presenting widely communicated visions for the integration of virtual environments into urban life. These initiatives offer valuable insights into the practical applications of the urban metaverse and showcase the range of new use cases emerging in the Smart City context. Building on these examples, we can explore the technological foundations and socio-economic design options that will shape the future development and adoption of the urban metaverse.

## Metaverse Seoul: A Virtual City-Platform for all Citizens

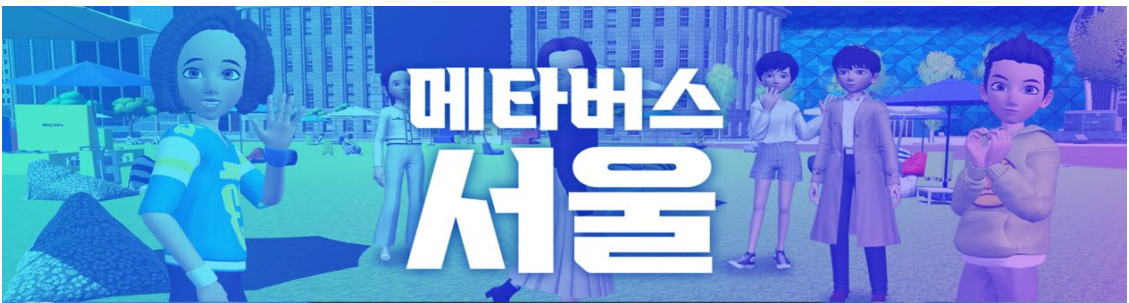

Source: https://english.seoul.go.kr/official-release-of-metaverse-seoul

A prominent and frequently cited example of an urban metaverse initiative is **Metaverse Seoul**, announced in 2021 and officially launched at the end of January 2023 following a beta phase. The project's core concept is to establish the metaverse as a virtual replica of public buildings and services on a single platform, making administrative and cultural experiences digitally accessible. As part of **Seoul Vision 2030**, the city plans to invest $38 billion into its development (Menzel, 2023).

The applications of Metaverse Seoul include citizen services, digital public engagement, and virtual tourism—essentially replicating the key functions found on most municipal websites but within an interactive, immersive environment (see box "Key Value Propositions"). Through the platform, citizens can access public services 24/7 and conduct official business in the virtual city hall, mayor's office, tax office, or Seoul Fintech Lab. Additionally, the popular AI chatbot **Seoul Talk** will serve as a virtual concierge, assisting users in navigating municipal services.

Beyond administrative functions, cultural and recreational spaces such as libraries, museums, parks, and Seoul's top ten tourist attractions are integrated into the metaverse. Even major public festivals, such as the New Year celebrations, will be digitally recreated within this virtual city. To enhance accessibility, VR headsets will be provided to citizens, enabling a fully immersive experience (Seoul Metropolitan Government, 2023).

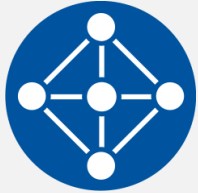

**Key value propositions of the Metaverse Seoul**

- **Networking:** Building a virtual world that offers immersive reality experiences and brings urban society together.

- **Free communication:** Creating a community space where users can express themselves creatively and communicate freely with each other.

- **Inclusion:** Interaction with avatars without barriers or discrimination, especially for young and physically impaired people.

According to experts, the primary motivation behind this initiative was the mayor's intent to establish a more direct communication channel with dissatisfied citizens, particularly in response to the sharp rise in rents and the ongoing housing shortage. The following excerpt from an interview (Weiss, 2022) underscores this point—Metaverse Seoul is not merely a technological vision but, above all, a social and urban cultural initiative aimed at fostering greater citizen engagement and public dialogue.

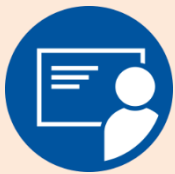

*"The biggest thing isn't a metaverse thing, but it's on citizen engagement (...) the real pivot when you move to citizen engagement is moving from a situation where it's essentially the city versus a problem, or oftentimes people versus the city versus a problem, to where it's people and the city versus a problem. (...) now, if you can actually create citizen engagement richly, it's actually the people and the city versus high housing prices.*

*Now it's a way too giant leap to say the metaverse is going to cure high housing prices, but it's not too far of a leap to say, 'This might be another episode or another way we can engage people with us as we face problems', as opposed to have them against us and then us against the problems."*

Mitch Weiss, Richard L. Menschel Professor of Management Practice, Harvard Business School

## Dubai Metaverse Strategy: Building a Global Industry

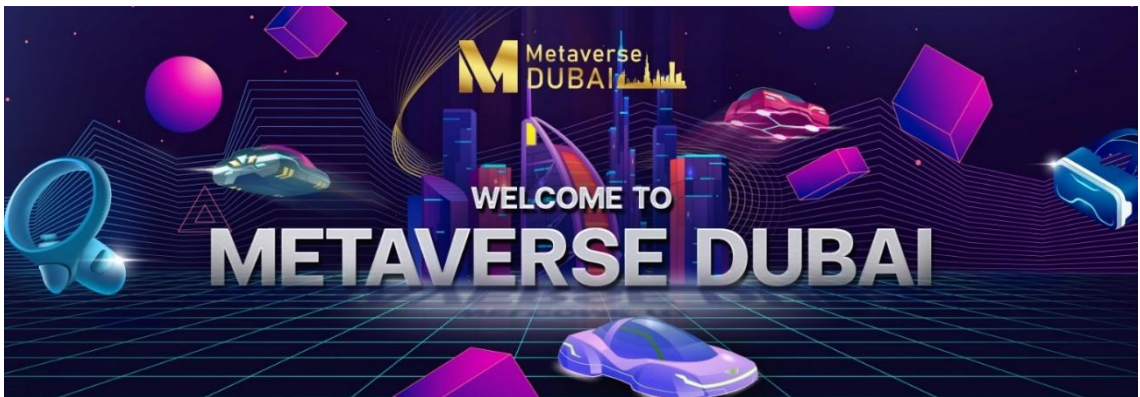

Source: https://metaversedubai.global

In contrast to Seoul, the Dubai Metaverse Strategy is primarily focused on economic development and the establishment of a new economic sector. The strategy aims to position Dubai as a global leader in the metaverse and Web3 industries, leveraging R&D collaborations and an **innovation ecosystem** built around startup incubators and accelerators. A key objective is to attract over 1,000 companies specializing in metaverse technologies, Web3, and blockchain, fostering an environment where international projects can thrive. By integrating the metaverse into its economic agenda, Dubai seeks to become a premier global technology hub, making the metaverse strategy a crucial pillar in the United Arab Emirates' broader digital transformation efforts.

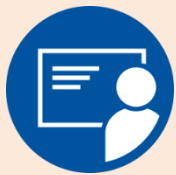

*"The metaverse will drive the UAE's and Dubai's efforts to provide innovative solutions, positively impact people's lives, and transform the city into one of the smartest hubs worldwide offering new economic opportunities. The Dubai Metaverse Strategy is in line with the objectives of the UAE AI Strategy to enhance the country's status as one of the world's leading countries in futuristic sectors by investing in new initiatives and empowering talent to drive digital transformation and the adoption of future technologies."*

Omar bin Sultan Al Olama, Minister of State for Artificial Intelligence & Digital Economy, United Arab Emirates

Dubai's metaverse Strategy also focuses on the development of **education and training programs** to build a talent pool for future digital professions. By fostering a skilled workforce, the city aims to create **new employment opportunities**, with a target of generating 40,000 virtual jobs by 2030 (United Arab Emirates' Government, 2023). Beyond workforce development, the strategy includes testing various metaverse applications and establishing global standards and regulations to support new government working models across key sectors, including tourism, education, retail, remote work, healthcare, and law. Additionally, public services are expected to be enhanced and streamlined through metaverse-driven innovations, improving efficiency, accessibility, and user experience in governance and municipal services.

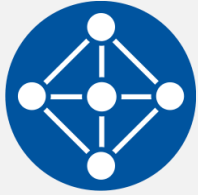

**Key Value Propositions of Metaverse Dubai**

- **Economic Development:** Building an international hub for metaverse technologies and applications, and in particular a talent base through education programs and incentives for foreign companies

- **Innovation:** Technology development and testing of applications for the Dubai Government.

- **Openness and Scaling:** Advocating for cross-city exchange to drive the adoption and scaling of secure platforms.

## Tampere Metaverse Vision 2024: A Model for An Inclusive Future

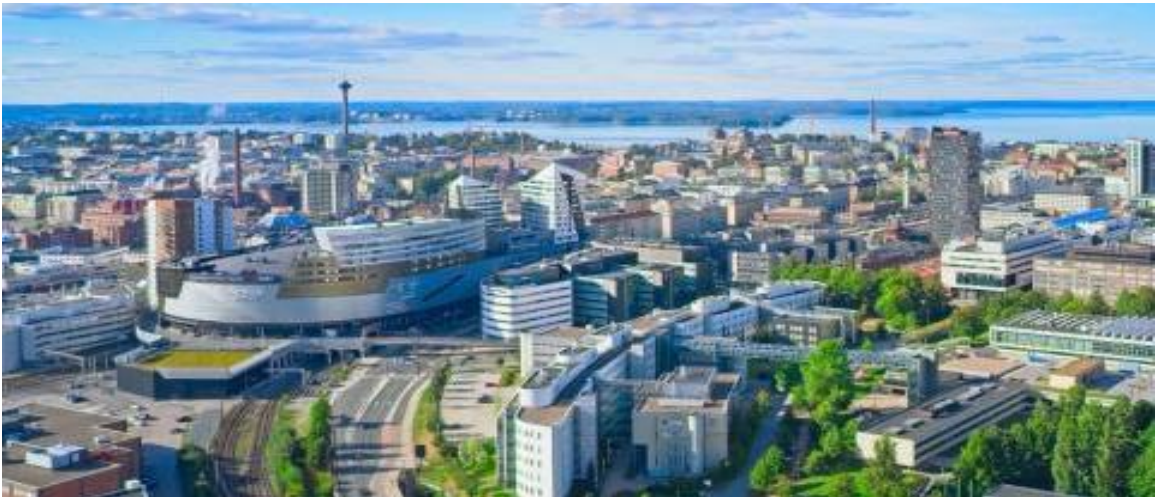

Source: https://www.tampere.fi

In 2023, the Finnish city of Tampere surprised experts by unveiling a sophisticated metaverse strategy for 2040. Developed in collaboration with the London-based Metaverse Institute, the strategy envisions the metaverse as an integral part of Tampere's urban infrastructure (Yan Zhang et al., 2023).

Known for its innovative mindset and technological leadership, Tampere aims to establish itself as a pioneer of the urban metaverse within the Scandinavian region. Unlike other metaverse strategies focused primarily on economic development, Tampere's vision positions the metaverse as a tool for urban planning and workforce development. The strategy provides a long-term perspective on how the city could evolve by the 2040s, offering insights into the future integration of digital urban environments.

Tampere's metaverse strategy is primarily focused on enhancing citizen participation, improving urban services, and optimizing functional processes. A key emphasis is placed on sustainability and inclusion, ensuring that all citizens can engage with the city's development, regardless of physical or socio-economic barriers. One major initiative is the virtual participation of Tampere's residents in city meetings and urban planning processes, enabling **broad and inclusive engagement**, particularly for those unable to attend in person. In the **education sector**, schools and institutions will integrate VR and AR technologies to create immersive learning environments. These include virtual classrooms, historical excursions, and interactive science labs, offering students innovative ways to explore knowledge. For tourism, Tampere has developed virtual tours, making historical landmarks, museums, and cultural sites digitally accessible. Through AR and VR, residents and visitors can experience cultural events and attractions in new, interactive ways, further strengthening Tampere's digital urban identity.

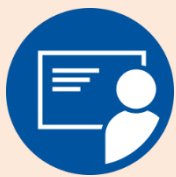

*"As a city organization, we need to be at the forefront of development. The vision provides guidelines for what could be possible in future Tampere. Our collaboration partners in the business sector, both existing and potential, also benefit from these metaverse guidelines. This is the direction we are headed towards through joint efforts and learning."*

Teppo Rantanen, Executive Director of Growth, Innovation & Competitiveness, City of Tampere

In **urban planning**, engineers and city planners leverage digital twins to simulate construction projects and analyze their impact on the city. These simulations help to identify potential planning errors early, optimize designs, and increase efficiency in infrastructure development. Additionally, IoT sensors are deployed throughout the city to collect real-time data on air quality, water levels, and other environmental parameters. This data is then visualized in the metaverse, allowing for interactive environmental monitoring. By integrating real-time insights, the city can proactively implement sustainability measures, enhance environmental protection efforts, and improve climate adaptation strategies—ensuring that urban development aligns with long-term ecological goals.

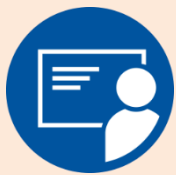

*"Our metaverse vision brings together a wide range of existing and emerging themes and concepts, including data utilization, digital twins, virtual realities, and artificial intelligence. These interconnected elements are driving the growing prominence of the urban metaverse as a transformative urban innovation. For us, the metaverse is not just a virtual space, but rather a continuously evolving digital environment that is seamlessly integrated into the real city. It serves as an extension of urban life, enhancing citizen interaction, urban planning, and service efficiency while bridging the gap between physical and digital realities."*

Tiia Joki, Development Manager, City of Tampere

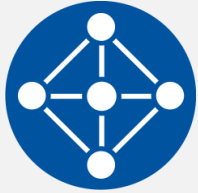

**Key Value Propositions of the Tampere Metaverse**

- **Increased Citizen Participation:** Utilizing virtual platforms to involve more citizens in decision-making processes.

- **Improved Educational Opportunities:** Innovative learning opportunities and a higher quality of education through VR and AR technologies.

- **Cultural Participation:** Virtual tours and events make cultural offerings accessible to a wider audience.

- **Efficient Urban Planning and Sustainability:** Digital twins improve planning processes, reduce costs and help with environmental monitoring.

## Zug Metaverse Pilot 2025: The Crypto Capital of Switzerland enters the Urban Metaverse

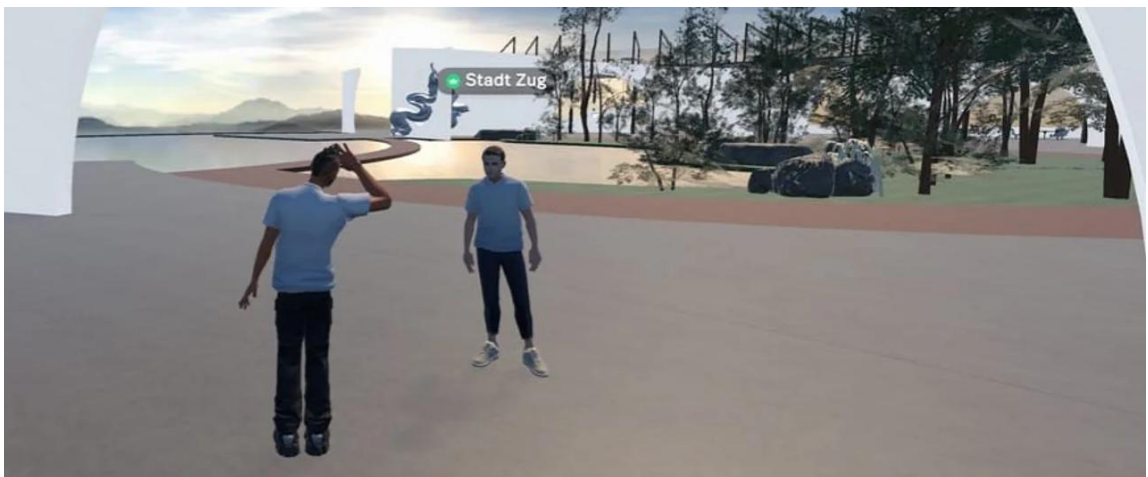

Source: https://www.stadtzug.ch

In March 2025, the Swiss city of Zug — internationally also known as the "Crypto Valley" for its leadership role in the adoption and support of crypto currencies — launched the country's first publicly accessible municipal metaverse. Built on the Spatial.io platform, the "Public Metaverse of the City of Zug" offers a digital space for interaction, cultural events, and civic engagement. The initiative aligns with Zug's broader digital strategy and reflects its ambition to integrate Web3 technologies into urban governance.

The metaverse prototype includes various virtual environments: an auditorium for hybrid events, digital art galleries, a rooftop bar, parks, meeting rooms, and scenic backdrops of Lake Zug. Citizens, local organizations, and businesses can rent these virtual spaces for meetings, exhibitions, or networking—adding a novel layer to public participation. Zug positions this digital realm as a future-oriented experiment in public innovation. The city's cultural department is using the metaverse for digital exhibitions and interactive cultural formats, aiming to make public art more accessible.

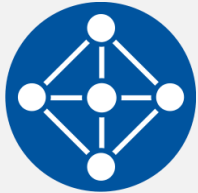

**Key value propositions of the Zug Metaverse**

- **Enhance Public Participation and Accessibility.** The Zug Metaverse aims to create an open, inclusive, and barrier-free digital space for civic interaction, enabling new forms of engagement between citizens, organizations, and the city administration.

- **Promote Art and Culture in a Virtual Environment.** By offering virtual galleries and cultural programming such as digital vernissages and art talks, the platform seeks to make Zug's public art collection and cultural activities more widely accessible and immersive.

- **Position Zug as a Leader in Digital Innovation.** As part of the city's broader digital strategy, the metaverse project rein-forces Zug's identity as a pioneer in emerging technologies, particularly in the context of Web3, blockchain, and virtual governance.

Despite its forward-thinking ambition, Zug's urban metaverse pilot has raised rather critical reports in the internal press after its launch about substance versus symbolism in digital innovation. The platform, while imaginative in concept, currently suffers from dated graphics, limited functionality, and low user engagement—casting doubt on whether it genuinely enriches urban life or serves primarily as a PR gesture.

Visitors report sparse interactivity and a sense of digital emptiness. Practical utility remains limited, especially when compared to more established tools for communication and participation. While the city promotes the platform as a model of accessibility and inclusion, the actual experience feels more like a prototype than a fully realized civic space. Moreover, concerns around digital privacy and the monetization of identity—such as avatars tied to external accounts or virtual goods—suggest a tension between innovation and user trust. For a city that accepts taxes in Bitcoin, this may seem like a natural evolution, but public digital spaces must be built with clarity around rights, risks, and responsibilities.

Ultimately, Zug's experiment is commendable for its vision, but its long-term success will depend on its ability to move beyond novelty and deliver genuine value to citizens in both form and function. The project serves as a timely reminder that building the urban metaverse is not just a technical challenge—but a civic one. However, without a start, even a bumpy one, no progress. And the sheer fact that the launch of a rather simple urban metaverse pilot still can generate over 50 press reports, including respectable national media like SZ, NZZ, and FAZ, is an indicator for the ongoing appeal of the metaverse.

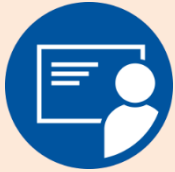

*"The City of Zug is tentatively venturing into the metaverse, arriving both too late and too early. Too late because the big metaverse hype was three years ago and has been replaced by AI in the public debate. And too early because the real metaverse may still be a decade away. The hardware required is complex, expensive, and therefore not widely available. Furthermore, it is still unclear who should do what.*

*The examples that have already been realized are always the same: virtual spaces for meetings, exhibitions, or concerts. Apart from the fact that the need for more virtual meetings has not exactly increased after a pandemic in lockdown, the metaverse is not a place but an interface—and with it comes the hope of enabling more complex and "physical" forms of interaction with networks, data, and services.*

*The first tentative steps taken by the city of Zug do not even begin to explore the potential of this abstract concept."*

Guido Berger, Director Digital of SRF (National Swiss Television), in a report about the Zug Metaverse initiative (Source: https://www.srf.ch/news/schweiz/zug-in-digitaler-parallelwelt-erste-schweizer-stadt-schafft-oeffentlich-zugaengliches-metaverse)

## Conclusion from the Visions of the Pioneers in the Urban Metaverse

Beyond **Seoul, Dubai, Tampere**, and **Zug**, several other cities are also pursuing Urban Metaverse initiatives, with a particular focus on the development of digital twins. These virtual models are being used to optimize urban planning, sustainability efforts, and citizen engagement.

**Singapore** has created a Digital Twin of the city to enhance urban planning and simulate different scenarios. This allows planners to assess the impact of new buildings on the urban climate and improve traffic flow and mobility solutions before real-world implementation. **Helsinki** is leveraging its digital twin to drive sustainable urban development. The Energy and Climate Atlas enables real-time monitoring and optimization of buildings' energy performance and consumption, while urban microclimate modeling helps identify heat islands and develop measures to improve air quality. **Amsterdam** is using a digital twin to enhance infrastructure planning and citizen participation. Through virtual tours, residents can engage with urban planning projects, explore upcoming developments, and contribute their feedback.

The examples of Seoul, Tampere, and Dubai illustrate that cities worldwide have high but distinct expectations for the Urban Metaverse. In Seoul, the metaverse is primarily designed as a direct communication and engagement platform for citizens, offering new ways to interact with public services and urban spaces. Tampere, in contrast, takes a long-term approach, positioning the metaverse as a tool for shaping the city's urban future leading up to 2040. Meanwhile, Dubai's approach is economically driven, with the metaverse serving as one pillar in its broader ambition to establish itself as a global technology and business hub for Web3 and digital industries.

These diverse strategies highlight the multifaceted potential of the Urban Metaverse, showing how it can serve as a platform for citizen engagement, sustainability, economic transformation, and future-oriented urban development.

Or, as the Dutch consultant and futurologist Harmen van Sprang puts it:

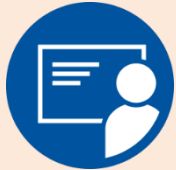

*"The metaverse is the ideal technology to empower a city and to make lives better in whatever that may mean."*

Harmen van Sprang, Co-founder & CEO, Sharing Cities Alliance & Studio Sentience, Amsterdam

Harmen van Sprang is one of the most well-known voices in discussions about the smart city of the future. However, he also emphasizes that, despite promising applications and significant investments, many critical questions remain regarding practical implementation, long-term financing, and citizen acceptance. Addressing these challenges will be essential for cities to ensure the long-term success of metaverse projects (Geraghty et al., 2022). To navigate these uncertainties, urban pioneers are needed to serve as models for other cities. While a fully developed urban metaverse does not yet exist, these pioneering cities provide valuable reference points that offer insights into what the future may hold. In the following sections, we will first examine the technological foundations of the metaverse in the smart city, followed by an exploration of the socio-economic design factors. Together, these technologies and design principles form the essential building blocks of the urban metaverse, shaping its development, functionality, and adoption.

## Technological Design Factors

The urban metaverse is built upon a range of key technologies (Aloqaily et al., 2023; Dincelli & Yayla, 2022; Zhang et al., 2024) that enable the creation of virtual worlds for design, simulation, management, and optimization of urban and community processes. These technologies form the foundation for enhanced urban planning, improved citizen engagement, and more efficient service delivery within smart cities. Figure 2 below provides an overview of these core technologies, which will be briefly introduced in the following section. After this initial overview, we will take a closer look at selected technologies that play a central role in shaping the metaverse within smart cities, highlighting their applications, challenges, and potential for urban transformation.

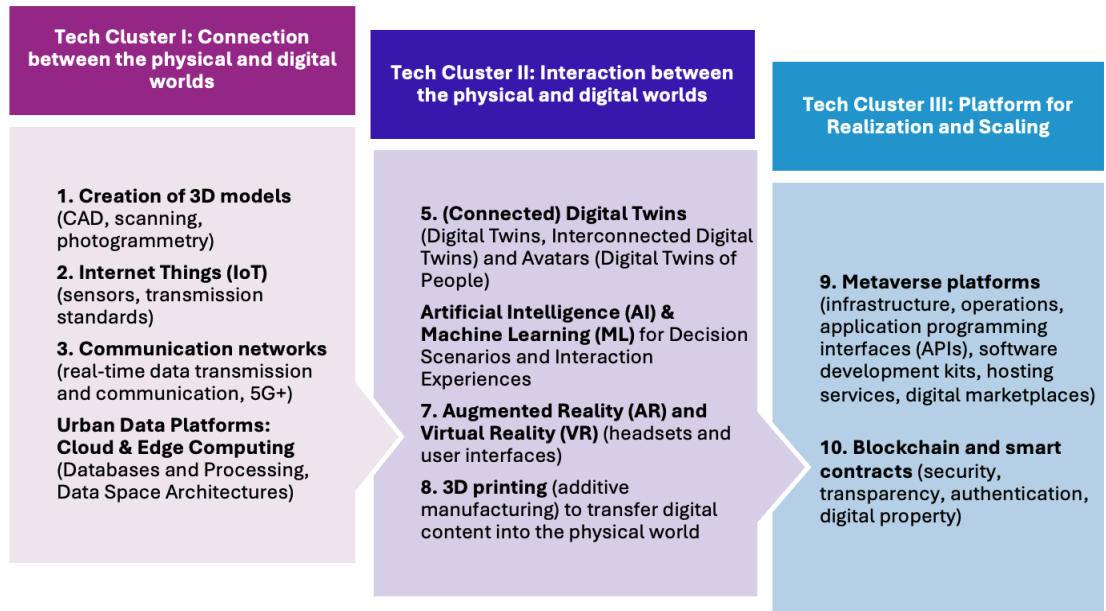

**Tech Cluster I: Connection between the physical and digital worlds**

**1. Creation of 3D models** (CAD, scanning, photogrammetry)

**2. Internet Things (IoT)** (sensors, transmission standards)

**3. Communication networks** (real-time data transmission and communication, 5G+)

**Urban Data Platforms: Cloud & Edge Computing** (Databases and Processing, Data Space Architectures)

**Tech Cluster II: Interaction between the physical and digital worlds**

**5. (Connected) Digital Twins** (Digital Twins, Interconnected Digital Twins) and Avatars (Digital Twins of People)

**Artificial Intelligence (AI) & Machine Learning (ML)** for Decision Scenarios and Interaction Experiences

**7. Augmented Reality (AR) and Virtual Reality (VR)** (headsets and user interfaces)

**8. 3D printing** (additive manufacturing) to transfer digital content into the physical world

**Tech Cluster III: Platform for Realization and Scaling**

**9. Metaverse platforms** (infrastructure, operations, application programming interfaces (APIs), software development kits, hosting services, digital marketplaces)

**10. Blockchain and smart contracts** (security, transparency, authentication, digital property)

**Figure 2: Three technology clusters of the Urban Metaverse**

## Fundamental Technologies of the Metaverse

The core idea of the metaverse is the connection between the physical and digital worlds. A key **technology cluster** enables the transfer of the physical world into the digital space and ensures a seamless link between both realms.

1. **Technologies for Generating 3D Models:** In addition to traditional 3D modeling in CAD environments, existing physical objects in urban areas are captured using 3D scanning or photogrammetry, the latter using photographs to generate detailed and realistic 3D models
.

2. **The Internet of Things (IoT):** IoT devices and sensors collect data from the physical world and exchange it with the urban metaverse to enable real-time monitoring, control, and optimization of municipal processes. Interoperability standards ensure that various devices, sensors, and systems work together seamlessly within the urban metaverse.

3. **5G and Future Communication Technologies:** Fast and reliable communication networks are essential for real-time data transmission and communication within the urban metaverse, particularly for digital twin applications that demand quick response times.

4. **Cloud Computing and Edge Computing for Urban Data Platforms:** These technologies provide the necessary computing power and data processing capacity to store, process, and access large amounts of data in real time, which is crucial for the scalability and efficiency of the urban metaverse.

In addition to these core technologies, a **second technology cluster** forms the foundation for interacting with the digitally connected world and bringing the digital world back into the physical realm.

5. **Digital Twins (DTs)** are virtual replicas of physical objects, systems, and processes that interact with each other and users within the urban metaverse. Enabled by IoT technology, they are always up-to-date. In the urban metaverse, digital twins are used to analyze processes, run simulations, and test improvements without affecting real-world systems. These digital twins exist at various levels, ranging from simple geographical data and infrastructure to individual buildings, machines, or components. The metaverse is characterized by the connection of multiple digital twins, creating more complex virtual worlds, known as interconnected digital twins (IDTs). Citizens and city employees can also have their digital twins represented as avatars.

6. **Artificial Intelligence (AI)** and **Machine Learning (ML)** can be used in urban metaverse applications to recognize patterns, make predictions, automate decisions and improve over time through experience. AI and ML support users in consolidating the multitude of data into meaningful decision scenarios and interaction experiences.

7. **Augmented Reality (AR)** and **Virtual Reality (VR)** are technologies that create immersive environments (based on connected digital twins) in which users can interact with digital content in a way that mimics or extends the physical world. VR creates a fully immersive environment, while AR integrates digital information into the real environment. While VR headsets still dominate as a user interface today, AR technologies will become increasingly important as an interface to the metaverse in the medium term.

8. **Digital Production Technologies (3D printing)** are less frequently mentioned in the context of the metaverse but can be seen as a means to directly transfer digital content into the physical world. Within the urban metaverse, developers and engineers can design and test complex products and components in a virtual environment, while 3D printing allows for the rapid transformation of these virtual designs into physical prototypes. This process can accelerate the innovation cycle by integrating design, manufacturing, and market feedback more quickly and efficiently. Additionally, 3D printing supports digital supply chains, enabling decentralized manufacturing that can, for example, produce spare parts on demand and close to the point of use.

The **third technology cluster** is platform technologies, which integrate the previously mentioned technologies and applications, providing the concrete application layer where urban metaverse concepts are realized.

9. **Metaverse Platforms** and **Urban Data Spaces** are a fundamental part of the ecosystem surrounding the urban metaverse. They provide the infrastructure that supports the interaction of various metaverse technologies. Metaverse platforms offer the necessary tools and interfaces for developers to create digital content and applications, including software development kits (SDKs) and application programming interfaces (APIs). These platforms also provide hosting services for content and applications, ensuring availability and scalability. Additionally, metaverse platforms feature marketplaces for buying, selling, and exchanging digital assets, as well as spaces and tools for social interaction, particularly for collaboration and community building, allowing users to meet, communicate, and work on joint projects.

10. **Blockchain and Smart Contracts** ensure a secure and transparent method for decentralized storage and transfer of data, as well as for the automation of contracts—programs running on the blockchain that execute automatically under predefined conditions. In the urban metaverse, blockchain and smart contracts are particularly useful for supply chain management, authentication, and copyright issues. NFTs (non-fungible tokens), a blockchain-based technology, are used to verify the uniqueness, ownership, and authenticity of digital and physical objects created with digital technologies.

These technologies form the backbone of the urban metaverse, playing a crucial role in the digital transformation of cities. In the following, we will delve deeper into what we consider to be the four central technologies of the urban metaverse: Urban Data Spaces, 3D Modeling, Networked Digital Twins, and Avatars and AR/VR technologies.

## Urban Data Spaces

The first of the four key metaverse technologies relevant to the urban metaverse is **urban data spaces**. Cities typically already possess a wealth of data. The first step toward a comprehensive digital solution involves breaking down data silos among the relevant urban stakeholders and either consolidating the existing municipal data into a central urban database or networking decentralized data spaces (Ritala, Ruokonen & Kostis, 2024; Clough & Wu, 2024).

Urban data spaces can thus be viewed as a specific form of metaverse platform, as previously discussed. An important development in this area is the shift from traditional **centralized databases**—structured systems designed for storing, querying, and managing data—to the concept of **decentralized data spaces**. While databases are typically optimized for internal use within an organization, focusing on consistency, integrity, and performance, data spaces create a collaborative and federated data environment where various stakeholders can share data dynamically without relying on a single central storage system (Redeker, Weskamp, Rossl & Pethig, 2021; Otto & Jarke, 2019).

A data space is constructed through the description of information technology relationships among various participants, formalized in a reference architecture. This framework establishes common standards and guidelines for storing and collaboratively using data within one or more ecosystems. Data spaces support the secure, controlled, and interoperable exchange of data between independent participants, often ensuring data sovereignty for the original data providers. A notable application of data spaces is the concept of **networked digital twins**, which will be further explored in the following section.

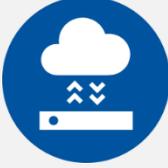

Example: Data Space
**The Mobility Data Space (MDS)**

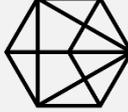

Mobility
Data Space
Data Sharing Community

- The Mobility Data Space (MDS) is a data marketplace where equal partners from the mobility sector can exchange data. The data provider retains ownership of the data at all times and has control over whether and with whom they share it. The goal is to create a cross-company data economy to foster the development of innovative, environmentally friendly, and user-centric mobility solutions.
- The MDS brings together companies, organizations, and institutions, including those looking to monetize their data and those seeking data for innovative mobility solutions. In many cases, this exchange creates a win-win situation for both sides. The MDS provides a trusted framework for such cooperation.
- Over 200 stakeholders from the German mobility landscape, including participants from science, industry, and public administration, collaborated on the conception of the MDS. The acatech Foundation managed and coordinated the process. In 2021, DRM Datenraum Mobilität GmbH, the sponsoring company of the MDS, was founded. It operates as a neutral non-profit organization with the task of further developing and orchestrating the Mobility Data Space both technically and commercially. The Mobility Data Space continues to receive funding from the Federal Ministry for Digital and Transport (BMDV).

Source: https://mobility-dataspace.eu/de/mobility-data-space

The general and case-specific added value of integrating and exchanging data is multifaceted:

- **Efficient, Open Collaboration**: The use of shared databases facilitates more efficient collaboration, especially in international projects and with interdisciplinary teams working across multiple locations. Previously, communication and collaboration often led to friction, as companies involved in construction projects used their own proprietary software within closed data silos. Data was exchanged manually between these silos, leading to double entries and inefficiencies. This resulted in a fragmented, lengthy planning and construction process where information could be lost, and planning collisions occurred, such as tearing up the same road multiple times. By integrating and exchanging data, construction measures can now be accelerated across the process chain from planning to implementation.

- **Improved Data Availability and Transparency**: With standardized, validated data, supported by AI or ML-based data profiling, the data is cleansed and exchanged automatically in real-time. The introduction of authorization levels ensures controlled access to security-critical data. Users receive daily updated, reliable access to the harmonized data they need, based on their assigned roles and rights. This enables temporary access for certain stakeholders, fostering collaboration and enhancing citizen participation. Moreover, non-security-critical open data can be made available to urban society, including start-ups and universities, allowing them to create new digital services and open additional value creation channels.

- **Simple Scalability of the Database**: The generic structure of the data room allows for easy integration of additional data sets. For example, data interfaces can be set up to federal portals, such as the Mobilithek, state portals like the NRW Energy Atlas, and other city portals. This integration ensures that higher levels access the consolidated data from lower levels.

The City of **Aachen** has already collected a wealth of geospatial data in central geodata portals, adhering to the EU INSPIRE Directive. In addition to data from municipal open data portals, Aachen also gathers IoT environmental and traffic data, which is stored in the Aachen Data Pool (AC-DatEP). While there are isolated interfaces between these data pools, there is not yet a single, overarching data space or exchange. One reason for this is that some projects initially focus only on data collection and provision. The following info box provides an example of an urban data space, the KomIT project.

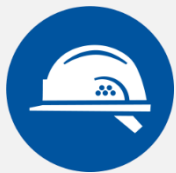

Example: Urban Data Spaces
**Project KomIT**

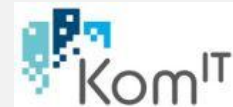

In the Aachen-based KomIT project, planning and asset data will initially be transferred to a central data room in order to redesign the planning and implementation of construction measures in an initial use case. Aachen has a wealth of planned and current construction projects involving many different entities - not only the municipal administration and its specialist departments, but also municipal companies such as the municipal utilities (STAWAG), network operators (Regionetz), IT infrastructure providers (regio iT) and transport companies (ASEAG).

Private companies such as planning offices and civil engineering companies are also regularly consulted throughout the planning and construction process. In addition, the City of Aachen endeavors to involve its citizens in change processes. A data room can help to optimize existing processes here.

Source: https://www.aachen.de/de/stadt_buerger/verkehr_strasse/verkehrskonzepte/ KomIT/index.html

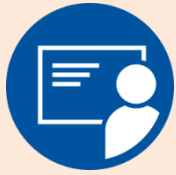

*"Structural change and the necessary redistribution of traffic areas will lead to many new planning and construction projects in the coming years. With the KomIT project, we want to overcome data silos and bring together municipal asset data such as supply and disposal lines, roads and street furniture in a common data space. In this way, we want to overcome today's usual information and time losses, planning collisions and, ultimately, considerable inconvenience for citizens."*

Dina Franzen-Paustenbach, Project Lead KomIT, regio iT gesellschaft für informationstechnologie mbh, Aachen

## Modelling of 3D Data

A city is like a living organism that can be viewed from various perspectives—ranging from a macroscopic lens, which shows the exchange of goods with other cities and countries, to a microscopic lens, which focuses on the movement of insects or individual dust particles (Lv, Xie, Li, Shamim Hossain & El Saddik, 2022). An essential requirement for the urban metaverse is a digitally mirrored city that accurately replicates these real-world details. This digital representation is based on partial digital models of the city, which serve as the foundation for a digital twin working with real-time data (see below).

The three-dimensional data and models, such as those representing streets, green spaces, buildings, bridges, lighting, street furniture, or charging points, are stored with their geo-references in an urban database or urban data rooms. Unlike conventional 2D georeferenced maps, **digital 3D models** provide a virtual and realistic depiction of cities, offering an immersive foundation for interaction within the metaverse.

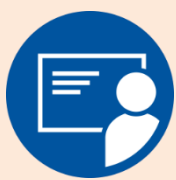

*"The recorded models and information (...) [bring] our city smart benefits such as the optimization of individual construction processes, information bases for maintenance and repairs, support in the area of space management and thus long-term cost savings and increased efficiency."*

Digitization Unit, City of Duisburg

These digital models can be created in various ways. In industry, digital models of objects have traditionally been developed in **CAD environments**, particularly for product development and design. More recently, these models are being increasingly supplemented by **game engines**, such as Unity or Unreal Engine, which provide extensive tools for creating interactive 3D objects and environments.

Artificial intelligence and machine learning are gradually automating and enhancing this process. For example, AI can assist in generating detailed 3D models from just a few data points, improving the realism and accuracy of existing models. **Generative AI** is emerging as a powerful content creation tool, allowing for the dynamic filling of the metaverse with both elementary and personalized content. However, high-quality content—such as the narratives that connect the content—will continue to be crafted by creative professionals, including engineers, artists, and academics (Hvitved et al., 2023).

In the smart city sector, as well as in many industrial metaverse applications, digital models are required for physical objects that have existed for years or even centuries, and for which no baseline digital data is available. Depending on the specific application or task, various te**chnologies can be employed to capture urban environments**. These technologies differ in terms of their resolution (and typically the size of the captured space) and their data collection frequency—ranging from one-off detailed images to regular recordings that generate time series.

For large areas, such as several city districts, **satellite remote sensing** is useful, with satellites equipped with different sensors to capture various spectral ranges, typically on a weekly basis. For higher resolutions, **airborne detection** using aircraft or drones can be considered.

Terrestrial methods include **mobile road scanning**, which many may have seen in everyday life through Google or Cyclomedia cars, as well as stationary laser scanning. These methods achieve the highest levels of detail, which are crucial, for example, when recording the condition of historical buildings for restoration purposes (Weckecker, 2024). **Laser scanning backpacks** and trolleys are also used by cartographers to capture interior spaces. For example, the Cathedral of **Aachen** has a website offering a 3D sightseeing flight of the interior.

Underground data, such as pipeline routes, is also essential, though often unknown in cities with a long history. This can be collected using **mobile ground radars**. Underground infrastructures change slowly or not at all, and they are often difficult to access. Therefore, as Lehtola et al. (2022) suggest, they should be monitored for maintenance and renovation needs after their initial mapping, for example, using IoT sensor technology.

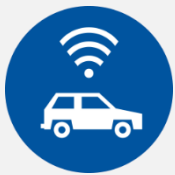

Example: 3D Capture of Streets in Aachen
**Mobile Road Inspection from Cyclomedia**

- As part of the Vista project, vehicles equipped with laser scanners from the German company Cyclomedia drove along and recorded the streets of the Aachen city region. License plates and faces were made unrecognizable to ensure data protection.

- The 3D model of the streets is used by the administration, for example, to offer citizens uncomplicated and quick building advice. Administrative staff can derive important information from the 3D model, such as the height of the surrounding houses or the width of the sidewalk, without having to physically inspect the building site.

Source: https://www1.wdr.de/lokalzeit/fernsehen/aachen/video-cyclomedia-macht-aufnahmen-der-staedteregion-100.html

As the names suggest, these detection systems typically operate with specialized **3D laser scanners**, known as LiDAR (Light Detection and Ranging). To perform the scan, the sensor emits laser beams in the infrared range and measures the time it takes for the beams to reflect off surrounding objects and return to the sensor. This process generates a point-based model of the environment, often represented as a black-and-white point cloud, sometimes enhanced with color images. These laser scanners are often paired with GNSS receivers to determine the sensor's position via satellite (HS Development & Services, 2023). Alternative methods include RADAR (Radio Detection and Ranging), which uses radio waves in a similar fashion, and photogrammetry, which mimics human depth perception by combining multiple 2D photographs taken from different perspectives to create a 3D model.

The development of mobile devices has further expanded this range of technologies, enabling **crowdsourcing** of user-generated content. Modern smartphones are equipped with LiDAR sensors, allowing users to easily virtualize objects and spaces in compatible applications (Hölzle et al., 2023). One example is Polycam, which allows users to generate 3D models across platforms[1]. In the future, citizens could contribute their generated 3D content to update the city's digital twin on a daily basis.

The collected images can then be integrated into a map. For instance, the city of Munich distinguishes between a detailed map and a fresh map. The detailed map focuses on achieving the greatest possible level of detail, suitable for unique buildings or historically significant structures. The fresh map, on the other hand, is continuously updated but accepts lower resolution, making it ideal for temporary events like the Oktoberfest (Bernhard, 2024). In Singapore, maps are refreshed every five years from the air and every two to three years in the streets (Poon, 2022). The **collection and synchronization of this data** should be designed with cost-benefit considerations in mind.

In the future, mobility systems may come standard with sensor technologies that can capture environmental data at short intervals. Autonomous systems, such as cars or drones, will also contribute to data generation & sharing (Lehtola et al., 2022). Firms like Nexar are already developing crowdsourced vision platforms, which process data such as images and videos using AI to generate maps with real-time content. [2]

## (Networked and Interconnected) Digital Twins

A core element of the industrial metaverse is **digital twins**, which are real-time representations of processes and objects, such as products and plants. These digital twins are synchronized with a specific frequency and accuracy throughout the entire lifecycle—from development and design to production or implementation, and continuing through the use phase. They support decision-making by integrating simulation models (Digital Twin Consortium, 2020). A digital twin serves its purpose when it is seen by decision-makers as fully equivalent to the physical asset, meaning that an experiment with the digital twin provides the same insights as an experiment with the physical asset.

---

[1] https://poly.cam/

[2] https://mapping.getnexar.com/

In the smart city context, digital twins bring 3D models of urban assets, such as infrastructure components, machines, or even entire facilities, to life and keep them current. This transforms pure representations into active planning objects. Digital twins access real-time and historical data, using this data in simulation models to enable more effective decision-making, optimize systems, or develop new products and services (Fukawa & Rindfleisch, 2023; Stark, 2022).

Until now, digital twins have mostly existed in functional silos. There is not yet a single, unified digital twin of the city, but rather a collection of specialist twins that are application-specific. These are often based on a geospatial twin that integrates all relevant geodata. One example is the mobility twin of a city (Deutscher Städtetag, 2023). In the industrial context, companies are extending this concept by linking digital twins and aggregating data and models from various production resources into a single data platform or data space, creating what is known as the networked digital twin.

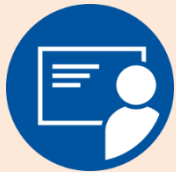

*"Germany's federal structure can be an obstacle to the implementation of an urban metaverse. There are currently many Gallic villages where a lot depends on individual players. The most important step is the structured networking of these players."*

André Henke, Program Management Digital Administration Lower Saxony (DVN), Lower Saxony Ministry of the Interior and Sport

The concept of **interconnected digital twins** (IDTs) refers to a network of virtual representations of physical objects or systems—a twin of the twins—where communication takes place in real time across their lifecycles. In this network, data is exchanged, and performance and decision-making are optimized through simulation models at the system level (van Dyck, Lüttgens, Piller & Brenk, 2023). The simulation models become more meaningful when digital twins from different actors are combined and used from various perspectives.

Deutsche Bahn is exploring this concept by developing a photorealistic digital twin of its rail network to plan and monitor operations. This allows for early identification of obstacles and more efficient use of existing resources without the need for additional tracks. As a result, train punctuality has improved, and the manufacturing and maintenance costs of infrastructure systems have been reduced. An AI-driven environmental perception system, using sensors on the front of the trains, will provide continuous data in the future (Deutsche Bahn, 2022; Geyer, 2022). In addition to the network infrastructure, research is also underway on digital twins of the trains and their components. For example, a twin of the entire drivetrain is shared with the train manufacturer, local maintenance center operators, and Deutsche Bahn's operations center.

Simultaneously, Deutsche Bahn is setting up a data platform for digital mobility twins, aiming to aggregate individual digital twins from its infrastructure and that of complementary companies, such as bus operators. The goal of these networked digital twins is to enable data-driven decision-making, like recommending the best means of transport to travelers. They also aim to improve performance across organizations and foster innovation in railroad infrastructure (DB Systel, 2020).

Networked digital twins are also central to the urban metaverse (Hartmann et al., 2023). The **urban twin of twins** brings together all the city's digital resources (Deutscher Städtetag, 2023). By integrating databases, existing services can be combined, improved, expanded, and linked, creating new services. This twin is accessible to multiple players within the smart city ecosystem, allowing complementary actors to use it with their data and models to develop new applications. Beyond the integration of city-level digital twins, there is also potential for further networking between the digital twins of cities themselves. This progressive integration will contribute to an expanding metaverse, aligning with the EU's vision of a European "Cityverse", briefly introduced at the start of this chapter.

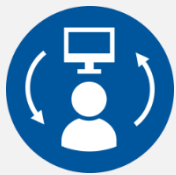

Example: Networked digital twins in the Smart City. **The Connected Urban Twin (CUT) from Hamburg, Leipzig and Munich**

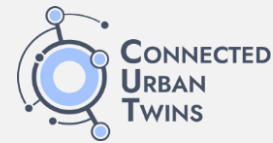

- DT initiatives from Hamburg, Munich and Leipzig have joined forces in the CUT project. The CUT project team sees a networked digital twin in the smart city environment as a construction kit with various building blocks. These building blocks can be put together in different ways depending on the requirements and circumstances of the respective city and include basic geographic information, specialist information, analysis tools and applications. The city's stakeholders are also part of the twin.

- The creation of a digital twin begins with the geodata that defines the object to be digitized. After the spatial delimitation, the specialized data, i.e. the application-specific data, is collected. This data helps to assign properties such as environmental data to specific locations. Apps are made available so that people can interact with data. Finally, a geodata twin is created to access the collected geodata and to integrate it with specialized data.

- Users can visualize and analyze the geodata twin in the app. In addition to the geodata twin, there are other twins that relate to different cases. However, the geodata twin is the central twin that coordinates the other twins and processes. The vision of CUT is that users can ask a question in a specific planning context. The application then accesses the twin and selects the necessary modules to answer the question.

Source: Hartmann et al. (2023); www.connectedurbantwins.de

Clear **standards and interfaces** are crucial for the successful networking of digital twins. A uniform language and a standardized data interface are essential for effective networking and the exchange of information. Standards enable seamless interoperability between different systems and participants, ensuring that data models are compatible and interchangeable via protocols, regardless of the platform or technology in use. Interoperability is a central element in both the industrial and urban metaverse, as various systems and devices must be connected to allow for smooth collaboration and data flow across different environments.

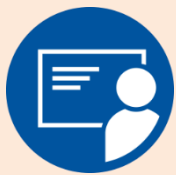

*"To turn the vision of the metaverse into a reality, firms can overcome the focality challenge by leveraging geometric and temporal affordances, and they can overcome the compatibility challenge through bilateral and multilateral cooperation. The path to the metaverse will require the arrival of consensus on shared standards for interoperability, which will allow existing proto-metaverses to interconnect."*

Andreas Kaplan, KLU Hamburg, und Michael Haenlein, ESCP Business School, Paris (Source: Kaplan & Haenlein, 2024)

In this context, the **Open USD standard** is gaining significant attention in the industry. USD (Universal Scene Description) is an open-source 3D format that allows for the representation, linking, and editing of digital models and digital twins across different software applications and source platforms. It is considered a key enabler for interoperability, particularly in the creation of complex digital models from various sub-models.

Originally developed by Pixar to efficiently display and manage 3D scenes in animated films and gaming, the flexibility and scalability of USD have made it widely adopted in other fields, including the development of smart cities. Today, a large number of companies have come together in the Open USD Alliance, aiming to establish the USD standard based on open-source principles. Prominent companies such as SIEMENS, with its Digital Xcelerator, and NVIDIA, with its Omniverse, are among those relying on this standard.

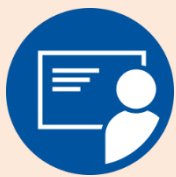

*"The aim is to make USD the industry standard or even the global standard. The metaverse will not work with proprietary data formats. A platform like this is only successful if it is as open as possible and every data supplier jumps on this format."*

Uwe Rechkemmer, Senior Specialist Simulation & Visualization, NVIDIA

In the urban metaverse, the USD standard is particularly valuable because it supports the creation of highly detailed 3D models. One of its key strengths is its ability to handle large and complex data sets. As smart cities involve numerous dynamic data sources—such as traffic flows, energy consumption, or weather conditions—USD provides a scalable solution to display and analyze these data in real time. This enables decision-makers to visualize future scenarios and assess the potential impact of urban development projects or infrastructure measures before their actual implementation.

## Avatars

Avatars are **digital representations** or characters that symbolize real people (or other living beings) within virtual environments. They enable users to navigate the metaverse, interact with it, and engage with other users (Hennig-Thurau et al., 2023). As digital representatives, avatars provide a visual and interactive presence in digital spaces, making them essential for fostering immersion and social connectedness in the metaverse (Davis, Murphy, Owens, Khazanchi & Zigurs, 2009; Nagendran et al., 2022).

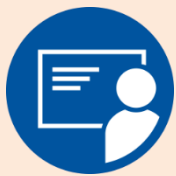

*"Analogous to the current state of the Internet with a multitude of websites, there will also be a multitude of interoperable metaverses. With the avatar, we can walk through these worlds, just as we switch websites in the browser today.*

*That's why everyone will have several avatars. For example, an avatar for dating or socializing, an avatar for work. By 2030, there could already be more avatars in digital worlds than people"*

Prof. Heiko von der Gracht, Chair of Futures Studies, Steinbeis Hochschule

Avatars can vary significantly in **appearance**, ranging from simple 2D icons to complex 3D models, and are often highly customizable, allowing users to design their digital selves to their liking. Specialized software solutions are available for creating avatars (Barta, Ibáñez-Sánchez, Orús & Flavián, 2024; Zhu & Yi, 2024), offering extensive options to personalize features such as physique, facial traits, skin color, clothing, and more. This enables users to craft personalized digital models of themselves. Additional methods, such as 3D scanning and the conversion of portrait photos using generative AI, provide further possibilities for creating realistic 3D avatars. For some individuals, especially on social media platforms like Instagram, presenting themselves digitally is important. The metaverse will offer even more opportunities for **self-expression** and creative self-realization (Hvitved et al., 2023). Fashion brands like Gucci, Dolce & Gabbana, and Nike are already designing virtual fashion today.

By collecting biometric data through wearables or even brain-computer interfaces (BCIs), avatars can evolve into **digital human twins**. This presents both opportunities, such as for medical treatments (Yan Zhang, Roberts, Kerimi, El Adl & You, 2023), and risks, such as the misuse of data for advertising purposes.

A recent development involves **AI-controlled avatars**, which do not represent specific individuals but act as representative personas of various stakeholders. These AI avatars can operate autonomously, analyze complex data, and generate personalized recommendations or warnings, thereby supporting decision-making and enhancing the efficiency of operational processes. In urban settings, AI avatars could function as virtual assistants for citizens, helping with information access, service navigation, and fostering community engagement.

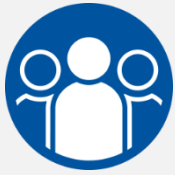

Example: Youth counseling with avatars in Seoul
**The Virtual Avatar Counseling Center**

At the Virtual Avatar Counseling Center of the Seoul City Council, young people can receive anonymous advice on various matters, such as their education and job, but also on relationship problems. In this way, barriers to participation in such services can be overcome, which might arise during face-to-face counseling.

Source: https://world.seoul.go.kr/policy/smart-city/metaverse-blockchain/

In the urban metaverse, avatars create new opportunities for citizen participation and urban management. Citizens can engage in virtual city council meetings as avatars, contribute to the design of public spaces, and participate in the planning of urban projects. These avatars can be linked to a unique **digital ID**, similar to a traditional ID card, enabling secure and personalized interactions within the metaverse.

## Augmented and Virtual Reality (AR & VR)

Augmented reality (AR) and virtual reality (VR)[3] technologies form the foundation for immersion and user interaction in the urban metaverse, allowing users to interact with digital twins or avatars in a way that mimics or extends the physical world. This enables high levels of participation, which is a key value proposition for many applications within the urban metaverse.

**Virtual reality (VR)** allows users to explore planning drafts for buildings, infrastructure, and other projects in an interactive, playful manner using VR glasses, while discussing these plans directly with urban planners in a virtual space—all without needing safety equipment or helmets.

On the other hand, **augmented reality (AR)** provides a technological alternative that overlays digital twins or other objects onto the real world, with just a tablet or smartphone as the end device. For example, the city of Bremen has developed its own AR app, BremenGo, which visualizes urban planning designs (Gellhaus, 2024) and brings tourist attractions, such as the Bremen Town Musicians, to life on an AR tour[4].

---

[3]  Virtual Reality und Augmented Reality werden zuweilen auch unter dem Begriff Extended Reality (XR) zusammengefasst.

[4]  https://www.bremen.de/bremen-go

For broader social participation, citizens can be invited to face-to-face events, such as discussing a construction project at a **digital planning table**. In the Digital Participation System (DIPAS) project, an interactive touch table displays the city in both 2D and 3D, and tells a story (in text form) based on the user's perspective. Citizens can provide direct feedback and learn about other participation formats. The software ensures a consistent experience across devices and generates statistics and key figures. Participating cities include Hamburg, Bremen, and Munich. Munich is currently working with SE3 Labs to make its digital twin more interactive using SpatialGPT, a voice-controlled AI, enabling users to engage with the city's digital model.[5]

## Socio-Economic Design Factors

To realize an urban metaverse and provide, exchange, and process the necessary data, a mature technological concept is not enough. It also requires interdisciplinary and cross-city cooperation among all stakeholders (Aloqaily, Bouachir, Karray, Al Ridhawi & Saddik, 2023). This section focuses on the socio-economic design factors of the urban metaverse.

The metaverse adds a virtual layer to physical cities and their organizational structures. Beyond the technological building blocks, there are several other important design factors that need to be considered. Just as in the physical world, intangible elements such as people's thoughts, visions, ideas, needs, fears, and wishes also influence this second reality. **An urban metaverse will only offer sufficient added value if it is used intensively by a sufficient number of relevant stakeholders over the long term** (Hvitved et al., 2023). This section outlines the design factors that can help avoid barriers to use and encourage participation in the urban metaverse.

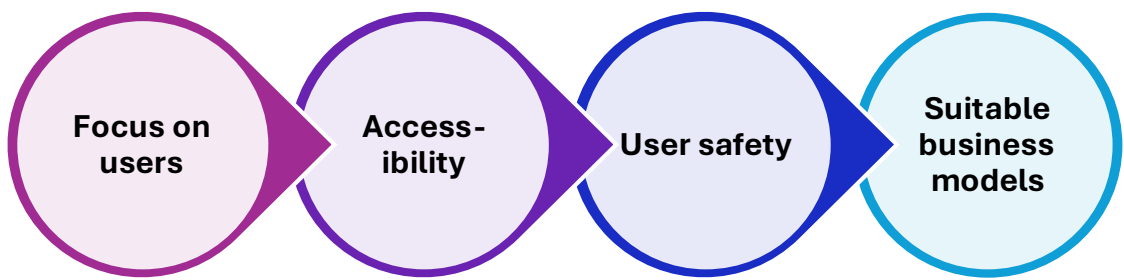

Figure 3: Four socio-economic design factors in the urban metaverse

---

[5] https://se3.ai/de/

## User Focus

To avoid disinterest or frustration during use, an urban metaverse must have **clearly defined and needs-oriented value propositions** for all users or stakeholders. Users' expectations regarding graphics, speed, interaction options, and similar aspects will be shaped by familiar experiences, such as computer games or films, as well as by media hype (Hölzle et al., 2023). A key strategy should therefore involve engaging future users early in the development process, embracing **open innovation and co-creation** (Guckenbiehl et al., 2021)—especially for complex, resource-intensive applications, such as those requiring advanced simulation capabilities and sophisticated tools for specialist users. In this context, lighthouse projects can play a crucial role in making the intended value more tangible, thus motivating more potential users to actively participate in the co-development of the urban metaverse (Hölzle et al., 2023; Thimm et al., 2024).

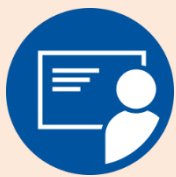

*"Applications in the smart city metaverse must solve existing problems and this benefit must be clearly communicated. I don't see that happening in many cities at the moment. It is not always possible to answer why this is actually being done."*

André Henke, Program Management Digital Administration Lower Saxony (DVN), Lower Saxony Ministry of the Interior and Sport

The **Outcome-Driven Innovation (ODI) method** offers a concrete approach that brings the **job-to-be-done (JTBD) philosophy** into practice. With the ODI method, specific "jobs" that users aim to complete are identified and then defined as concrete, measurable "outcomes". We have successfully applied the ODI method in the design of immersive e-learning environments (Piller et al., 2017) and the development of smart products (Hankammer et al., 2019). We also see significant potential in using job-based thinking to develop user-centered urban metaverse applications.

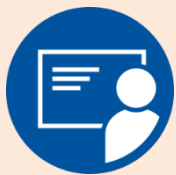

*"Jobs-to-be-done theory provides a framework for (i) categorizing, defining, capturing, and organizing all your customer's needs, and (ii) tying customer-defined performance metrics (in the form of desired outcome statements) to the job-to-be-done."*

Anthony Ulwick, Founder of the strategy and innovation consultancy Strategyn

To create a metaverse application for smart cities, the following principles of Outcome-Driven Innovation (ODI) can be applied:

1. **Identifying the JTBD (Jobs-To-Be-Done):** This step involves understanding what tasks users-such as citizens, urban planners, businesses, etc.-want to accomplish. It focuses on identifying basic needs, problems, or challenges in urban life that could be solved more effectively using metaverse technologies. This process typically involves in-depth interviews, and valuable secondary data from sources like social media and internet forums can also be useful. Here are some examples of potential jobs for different stakeholders:

- *Citizens:* More efficient access to public services, improving traffic flow, or creating community spaces in the virtual world.
- *City planners:* Visualizing future construction projects, simulating traffic scenarios, or analyzing the impact on the environment and infrastructure.
- *Companies:* Virtual presentation of services, simulation of business processes in the city, or interaction with virtual communities.

2. **Prioritize the JTBD and Derive Desired Results (Outcomes):** Once the users' jobs have been identified and formulated as concrete outcomes, they must be prioritized. It is crucial to focus on the (latent) needs that are most important from the user's perspective in urban metaverse development. A quantitative survey among a representative sample of users can help determine which outcomes are most critical and how well these outcomes are currently being met. The goal is to maximize user satisfaction by addressing the most important unmet needs, which is where the urban metaverse application should focus.

3. **Creation of Functionalities and Features Based on the JTBD and Outcomes:** With the identified jobs and outcomes as the foundation, the metaverse application can be designed to meet the specific needs of the users. It's important to also consider traditional technologies or services outside the urban metaverse that might solve the same problem more effectively. The urban metaverse should be not implemented for its own sake but needs to create value by solving open jobs of various stakeholders in the best possible way.

Examples of how the open jobs could be solved by metaverse applications include:

- ***Virtual representations of urban assets***: This addresses the need for citizens and city planners to experience and evaluate construction and environmental protection projects, such as new transportation routes, in an immersive environment.
- ***Interactive dashboards and simulations:*** These address the inadequacy of existing monitoring and planning tools by offering real-time data and predictive analytics that can optimize the use of urban services and guide forward-looking urban management decisions.
- ***Personalized user experiences through customized metaverse applications:*** This satisfies the need for tailored information and functionalities for different user groups.

By applying the ODI method and job-based thinking, the specific needs and tasks of users can be placed at the center of the design of applications for the urban metaverse. This ensures that the applications are not only technologically advanced but also user-centered, improving efficiency, collaboration, and quality of life in the city.

Urban metaverse applications hold the potential to serve basic social needs. For many citizens, participating in the urban metaverse as a place to meet and collaborate could already provide valuable added value. This aligns with the vision of big tech companies, which aim to connect leisure and work with the metaverse (Rostami & Maier, 2022). ARand VR provide playful elements that further enhance the attractiveness of urban metaverse applications (Hölzle et al., 2023; Rostami & Maier, 2022).

The metaverse can make participatory processes in cities more inclusive, effective, and transparent-from idea development and discussions to consensus-building and implementation. It can also create spaces for collaboration, such as a Citizen Collaboration and Co-Creation Hub, where people can collaborate on new urban solutions. In this way, municipal and inter-municipal innovation efforts can be supported and improved. In general, the opportunity for greater community involvement not only increases acceptance for the solutions created but also enables better-informed decision-making.

## Accessibility

A fundamental prerequisite for the broad use of an urban metaverse is that it must be **low-threshold and ubiquitously accessible** (Datta, 2022b; Hölzle et al., 2023; Hvitved et al., 2023). This is one of the three essential characteristics we mentioned at the beginning as the goal of the urban metaverse. An urban metaverse has the potential to bridge the digital divide, but it can also widen it (Hvitved et al., 2023). For instance, it could help non-mobile people overcome social isolation and maintain contacts more easily (Baker et al., 2019; Datta, 2022a). On the other hand, people in rural areas could be excluded due to poor internet connections (Datta, 2022b). Work-related limitations, such as IT security requirements, must also be considered, as they can hinder uncomplicated participation from business and government representatives. In addition, some user groups might be deterred by perceived risks (Hvitved et al., 2023).

It is crucial to examine urban society from multiple perspectives: **Where are there financial, professional, educational, geographical, and physical differences?** Only by addressing these questions can we ensure that people can participate in all digital social processes without barriers or discrimination. **Devices** like AR and VR glasses act as portals to the metaverse for users. Widespread availability of these end devices is essential to catalyze the use of the metaverse (Hvitved et al., 2023). Just as there are clothes for all types of people (e.g., children, adults, pregnant women), research should also focus on designing **human-centered technology** for end devices. This will be crucial to enable broad social participation. To make the metaverse truly accessible, international standards like the Web Content Accessibility Guidelines should be applied and further developed (Zallio & Clarkson, 2022).

Currently available glasses often lack ergonomic design and are generally made for adults (Hölzle et al., 2023; Zallio & Clarkson, 2022). Furthermore, facial expressions and gestures are only captured in newer models, which limits the ability to express oneself fully in the metaverse. Technological options that could support **physical immersion**, such as integrating haptics, olfaction, and temperature into the virtual experience, are still underdeveloped. For instance, several companies, including Meta, are working on haptic clothing, such as gloves.

Brain-Computer Interfaces (BCIs) could provide a solution here (Kuru, 2023; Yan Zhang et al., 2023), establishing a direct connection between the computer and the brain to generate sensory stimuli. The rise of BCIs raises ethical concerns regarding "mind reading" and potential manipulation, but it also offers significant opportunities, particularly for people with special physical conditions (Kaplan & Haenlein, 2024). Cities should stay informed about current developments and carefully weigh the pros and cons.

Immersion is not only physical—**mental immersion** is also crucial, as seen in gaming or reading a book, where one becomes absorbed in another world. To facilitate this type of immersion, many people, especially non-digital or non-virtual natives, need to be introduced to the metaverse, particularly in its early stages (Garrido & Gonzàlez, 2023). For younger generations who have grown up with smartphones and laptops, or for gamers who are already used to meeting friends in the 3D worlds of online games, the boundary between physical and digital reality is less distinct than for older generations (Allam et al., 2022; Hvitved et al., 2023; Kuru, 2023). Cities can accelerate this process by offering workshops and addressing the metaverse in schools (Hvitved et al., 2023), but above all, by providing appropriate training for all population groups.

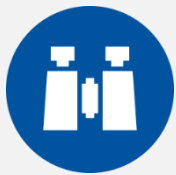

A Look Into the Future
**Which Technology Trends Increase Accessibility to the Metaverse for All Target Groups**

- *Wearability:* AR and VR technologies are increasingly being integrated into wearable devices such as smart glasses and headsets. This area of development is still in its infancy, and consumer complaints about discomfort with the new Apple Vision Pro after its launch show that there is still plenty of room for further VR improvements and innovations in the coming years.

- *Optics* are the most important part of an AR/VR device and also the most technologically demanding component, which prevents the devices from becoming cheaper. The optics in a headset define the field of view, resolution and eye comfort. Holographic optical elements (HOE) are currently an important technology for realizing complex optical functions due to their thin and lightweight design and the versatility of optical applications. Originally, HOEs were developed as a by-product of research in the fields of laser technology and holography. HOEs differ from conventional optical devices in that they bend and thus manipulate the light depending on the wavelength and angle of incidence

- *Integration of AI:* Advances in AI development have led to significant improvements in AR and VR technologies and improved rendering, tracking and processing. In particular, this has increased the realism of 3D characters and environments and impacted the dynamic possibilities of gaming scenarios and interactivity. This makes AI an important technology for the gaming sector.

- ▪ *Cross-Platform Applications*: Through WebAR and WebVR, users can immerse themselves in augmented and virtual reality without having to install special applications on their device. This is a major step towards making AR and VR accessible to a larger number of users on all platforms and operating systems. This promotes compatibility and a smoother, more consistent user experience.

Source: https://medium.com/@Web3comVC/the-near-future-of-vr-ar-f15d6d6a889b

In order to involve people with limited financial resources, one simple option could be to build VR glasses using a smartphone and foldable cardboard. However, such frugal innovations often come with compromises. For example, advanced functions like gesture and facial expression detection would not be available (Allam et al., 2022; Marabelli, 2023).

For cities, the procurement of end devices should be tailored to the use case and user group. It may be more practical to procure high-quality devices for a limited number of people (e.g., for specialist staff) or to acquire inexpensive devices for a larger group of people, depending on the specific needs of the application and the scale of participation.

## User Safety

The urban metaverse further emphasizes the social aspect of the rather technology-centric solutions typical for past smart city and urban digital twin initiatives: Besides technology, citizens and social life need to take center stage (Allam et al., 2022; Kuru, 2023). As a result, the urban metaverse is also becoming a **playing field for political and private sector interests**. Here, it must be ensured that the legal framework and governance of the metaverse platform are geared towards the well-being of citizens (Allam et al., 2022).

The metaverse offers unprecedented potential to collect **personal data**: physical data such as body size, gender and eye color, physiological data such as heart rate and perhaps even brain activity and behavioral data such as body movements, location history, interactions and transactions. The processing and interpretation of this data can allow profound conclusions to be drawn about the people using it. This not only enables the development of advanced authentication mechanisms, for example, but also the interpretation of facial expressions and emotions.

**Data gains context like never before** when it is collected in the environment of the metaverse (Marabelli, 2023). This offers many opportunities, but also creates new challenges. Even if there are still unresolved issues surrounding the storage and analysis of this data, often in real time, some of the opportunities, risks and challenges around user data are already obvious today (WEF, 2023):

- **Tracking, Tracing and Predicting Human Behavior:** A person who dives into the metaverse is represented there with their data and thus makes themselves vulnerable to **surveillance** (Allam et al., 2022; Hvitved et al., 2023). Employees could be fully monitored in industries characterized by performance pressure, citizens could be checked for their health and risk behavior by insurance companies, but users could also be observed for scientific field studies (Marabelli, 2023). Law enforcement agencies could also use metaverse data for their own purposes. In **predictive policing**, retrograde crime data is placed in the context of the city (and its data) to predict human behavior (Zallio & Clarkson, 2022; Marabelli, 2023). [6] Residential burglaries and vehicle thefts, for example, can be viewed in the context of the respective neighborhood, the time of day, the socio-economic background of individuals and groups, and the traffic situation. It becomes tricky when the **results** of such statistical evaluation methods are **distorted** by the selection and processing of the data. This can lead to groups and individuals being disadvantaged and discriminated against due to a qualitatively poor and impaired database (Muzaffar, La Torre & Gstrein, 2020).

- **Personalization of Content:** The shift of social life to the metaverse envisaged by large tech companies means that the individualization and personalization of content - such as the virtual living room or avatar - is becoming very important (Piller & Euchner, 2024). But what about users who don't want to reveal much about themselves? While it is possible on traditional social media platforms, for example, to simply not upload a profile picture, **anonymizing** avatars is more challenging (Arunov & Scholz, 2023). Generative AI, which produces photo-realistic avatars, can also be misused for **deep fakes**. Words could be put into the mouths of users and behavior attributed to them for which they are not responsible, making them vulnerable to defamation and manipulation (Interpol, 2024). The metaverse could also be a breeding ground for **personalized, behavior-based advertising** (bespoke behavioral advertising), through which users are targeted very specifically on the basis of their personal data (Zallio & Clarkson, 2022).

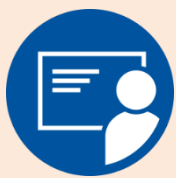

*"The early development stage of the metaverse presents a unique opportunity to prioritize privacy and safety by design, responsible innovation and human-first principles. While the metaverse is still in the early stages of development, stakeholders have a brief window of opportunity to build a digital world that embodies inclusivity, accessibility, equity, diversity and sustainability."*

World Economic Forum, Metaverse Privacy & Safety Report

In the urban metaverse, it is all the more important to establish standards and regulations for the **protection of personal data**, for example by setting clear guidelines for data management (Arunov & Scholz, 2023; Hvitved et al., 2023; Vessali, Galal, Nowson & Chakhtoura, 2022). While some users will willingly disclose their personal information in an urban metaverse, others are likely to be more reticent or even wish to remain anonymous. It should be possible to delete the digital footprint at any time.

---

[6]  Crime forecasting software is already being used in the USA, Europe and Germany, for example.

However, it is not only the misuse of personal data that can pose a risk; people themselves are also susceptible to **mental and physical harm** when using the metaverse. We are already seeing how addictive online games and social media can be. The metaverse also offers great potential to escape reality due to its immersive nature (Yan Zhang et al., 2023). For some, the metaverse will also have an overstimulating effect and lead to psychological stress (Zallio & Clarkson, 2022), especially if companies use the platform for intrusive advertising. Immersion in a virtual world also reduces awareness of the physical world, which can lead to loss of orientation and injury.

As with other interaction platforms, a safe user experience should be ensured. **Experiences and feelings in the metaverse are real** - so the immersive experience should not become a nightmare due to assaults such as stalking, bullying and sexual harassment. Clear identification mechanisms can help prevent perpetrators from hiding anonymously behind an avatar (Hvitved et al., 2023). Implementing a minimum distance between avatars and employing moderators can also be useful (Zallio & Clarkson, 2022).

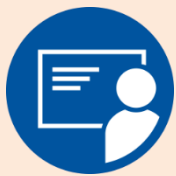

*"Privacy and safety in the metaverse will transcend the physical and digital worlds. Trust and well-being in the digital and physical worlds will continue to merge. Businesses, governments, academia and civil society should proactively collaborate to build and support appropriate standards and policies that support metaverse safety and privacy and take a human-first approach enmeshed in human rights considerations while furthering needed innovation in this burgeoning field. Ultimately a successful metaverse will be one that promotes trust, well-being, privacy and safety."*

World Economic Forum, Metaverse Privacy & Safety Report

**The urban metaverse is not a legal vacuum**. First of all, it is important to stipulate in user agreements which national and international law applies (Marko et al., 2023). [7] Existing regulatory requirements such as the General Data Protection Regulation (GDPR), but also labor law, commercial law and copyright law must be checked for their applicability in the context of the urban metaverse. There are still many legal questions here. In order to avoid an uncertain, dystopian metaverse, it is important to close legal loopholes, but also to find the right balance between over- and under-regulation. As the metaverse is based on many existing technologies, it may be possible to fall back on existing legal solutions (Hölzle et al., 2023). Cities should stay up-to-date here by actively following and helping to shape the amendment of existing regulations and laws or the adoption of new ones.

---

[7] The physical location of the users and platform operators is often decisive here.

Scope in the legal framework should be used by cities (or the respective central platform operator) to implement adequate **governance mechanisms** in the urban metaverse. However, it will be challenging to ensure a safer, non-discriminatory approach for urban society. Social media platforms have not yet found a solution for holistically moderating inappropriate content such as fake news and hate speech. This problem is exacerbated in the metaverse by the diversity of data, content and possible criminal offenses.

## Business Model Patterns

The development, implementation, ongoing operation and continuous development of an urban metaverse require both significant **initial investment** and a **continuous budget** in order to adapt the technology, content and the associated user experience to changing requirements and opportunities.

If these costs are not covered by tax revenue or public funding, it is necessary to clearly communicate the benefits for potential investors and to think about viable business models in order to ensure long-term financing (Endres et al., 2024; Mancuso et al., 2023; Latino et al., 2024). Corresponding business models should be systematically developed at an early stage, i.e. in parallel with the conceptualization of the urban metaverse.

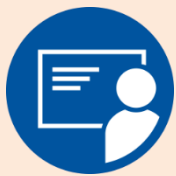

*"Everything is networked in the Smart City Metaverse. This networking of administration, municipal utilities, citizens, companies and all other stakeholders results in new business models."*

André Henke, Program Management Digital Administration Lower Saxony (DVN), Lower Saxony Ministry of the Interior and Sport

For a systematic exploration and creation of business models, we can draw on the many established approaches from the literature on business model innovation. [8]  Like all business models, business models for the urban metaverse are based on a **combination of innovative elements and proven patterns**, tailored to the networked physical and virtual environment. Important business model patterns for the urban metaverse **in the area of monetization and financing of an urban metaverse** are (not exhaustive) (Endres et al., 2024; Mancuso et al., 2023; Latino et al., 2024):

1. **Experiential Marketing and Virtual Advertising Space:** Companies can rent virtual advertising space such as advertising pillars and billboards in the Urban Metaverse in order to address citizens or city visitors directly in the virtual city environment.

---

[8]  See https://businessmodelnavigator.com/ for a list of business model samples from the BMI Lab St. Gallen.

Innovative advertising concepts such as the temporary design of the city environment in the brand design or extensive personalization are also possible. Brands can also offer sponsored events and experiences in the metaverse. Ultimately, advertising financing and experiential marketing are conceivable for all forms of entertainment in the urban metaverse. It is important to find the right balance in order to generate sufficient revenue and not scare users away with overly intrusive advertising.

2. **Marketplaces:** In the urban metaverse, there can be various e-commerce models for buying, selling and renting the previously advertised digital and physical products. The same is conceivable for the provision of services. Local retailers can also present and sell physical products in virtual stores, showrooms and malls. Customers can interact virtually before purchasing the products in the physical world. As is usual with regular online marketplaces such as Amazon or eBay (and also classic physical weekly markets), the operators of the marketplace charge a fixed or transaction-based fee for all virtual sales.

3. **Token-Based Economy:** Virtual currencies and non-fungible tokens (NFTs) enable the ownership and trading of digital assets such as real estate, creating new markets in the metaverse. For example, users can rent virtual real estate such as premises or buildings that they use for social, business or cultural purposes.

4. **Monetization of Digital Twins:** Digital twins of urban infra-structures enable precise simulations and predictive analysis. Monetization is achieved through the licensing of real-time data (or regularly updated models) used by public utilities, planning offices or companies. Providers of the digital twins can be the city itself, but also contracted or independent private service providers who are able to create the digital twins in real time by accessing the corresponding physical assets (infrastructures).

5. **Monetization of Urban Data:** The central idea of the urban metaverse is the constant updating and updating of the digital world in line with the development of the physical world. This requires operational real-time data, e.g. traffic flows, pedestrian frequencies, climate data, etc. These complement the digital twins in the creation of predictions or prescriptions. Providers can charge a fee for the secure and data protection-compliant provision of this data (as mentioned in the last section). Data subscriptions can support the regular use of digital city models for urban planners, architects or infrastructure operators to simulate or optimize projects.

6. **Monetization of Content and Services**: User-generated content and commercial applications can also be created in the urban metaverse and made available to other users, city administrations, engineering firms, real estate companies, etc. via micro-transactions. For example, event organizers can hold virtual concerts and festivals and charge admission, architectural firms can offer their virtual building designs, and companies can offer special applications with extended planning, simulation and analysis options for digital twins.

7. **Freemium Models** are a special case in which free basic services are supplemented by premium offers. For example, users can receive extended access to data, content and personalized services. This is particularly interesting for companies and public authorities.

8. **Taxes and Fees** could also be an obvious source of funding for cities and municipalities to finance an urban metaverse. As long as the virtual world is still close to the physical world, i.e. companies still have a clearly definable domicile and services still have definable recipients, the easier it is to transfer the current tax legislation to the metaverse. However, the further the virtual world moves away from the physical world and the more complex the metaverse of different cities merges into a "cityverse" (the metaverse of urban metaverses), the more the current law reaches its limits (Kim, 2023; Pandey & Gilmour, 2024).

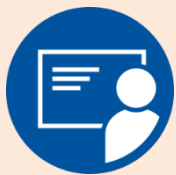

*"If you have big numbers of users, it's no problem to have a business model in place."*

Harmen van Sprang, Co-founder & CEO, Sharing Cities Alliance & Studio Sentience

**The metaverse spans the framework around digital resources and technologies** such as blockchain, 3D modeling and machine learning for data analysis. Their combination results in new fields of application and **different starting points for business models** (Hölzle et al., 2023). This makes the metaverse a promising field for scalable business models. The business model patterns presented above can be used to make the virtual space of the urban metaverse commercially viable. Among other things, the revenues can refinance the maintenance and expansion of (real) urban infrastructure.

**Platforms** are not only a central technological building block of the metaverse, they are also the basis for the business model patterns described above. Platforms in the urban metaverse mediate between urban players and create **ecosystems of shared value creation**. Ecosystems bring developers together and create digital content and services, while virtual marketplaces enable them to be traded. For further reference points, we refer to the literature on platform business models, which now fills entire shelves.

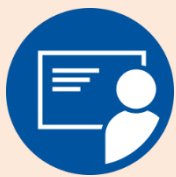

*"The provider landscape is currently very fragmented. Small-scale, regional development can be a great strength."*

Prof. Heiko von der Gracht, Chair of Futures Studies, Steinbeis Hochschule Berlin

The systematic development of digital business models and the orchestration of platform-based ecosystems is often not a core competence of municipal administrations. Dedicated consulting firms such as regio iT, which are just as familiar with the municipal environment as they are with the latest developments of leading digital companies, can make an important contribution here. They can bring together the necessary authorities and stakeholders such as municipal companies, schools and universities. They can also help to further utilize and market self-developed applications, for example by making them available to other cities and municipalities in the Urban Metaverse.

## Systemic Challenges for the Implementation of an Urban Metaverse

Like all technological infrastructures, a state-of-the-art urban metaverse will be overtaken by further technological developments over time (Arunov & Scholz, 2023). Increasing demand and adaptation by new user groups may also require further adaptation. Therefore, **"flexibility by design"** should already be followed when designing the infrastructure so that it can be flexibly upgraded over the lifecycle, for example through a modular architecture. It is important to work with recognized standards and protocols in order to ensure the interoperability of the individual modules, such as the IoT sensors and devices. The same applies to the scalability of the business model patterns used.

In an urban metaverse, it is therefore crucial from both a technical and socio-economic point of view to make **data available across all areas and sectors- but at the same time fairly and securely** -- in order to realize a comprehensive representation of the complex overall city system (Guckenbiehl et al., 2021). To achieve this, data silos must be broken down and blind spots eliminated through appropriate measures such as the installation of IoT sensor technology. Data quality and uniform data standards are particularly important here (Vessali et al., 2022).

However, with current **network technologies** and still insufficient network coverage in some places, the metaverse will not yet be fully realized, as the enormous amounts of heterogeneous data cannot be transmitted (Hvitved et al., 2023). Technologies such as 5G and 6G can be an enabler to ensure sufficient transmission capacity with low latency (Aloqaily et al., 2023; Kuru, 2023). This is the only way to render 3D environments in detail and guarantee a smooth, real-time experience for all participants without causing cybersickness.

Processing data at the level of a fully networked city in the metaverse still requires enormous computing power, which according to Intel exceeds the **computing power** available today by a factor of 1,000 (Koduri, 2021). Since the computing power of chips will not increase sufficiently in the coming years according to Moore's Law, high-performance technologies such as quantum computing may be required (Yan Zhang et al., 2023). Edge computing also offers the opportunity to process data decentrally at the "network edge," thereby reducing the load on networks (Hölzle et al., 2023).

Finally, the urban metaverse with its energy-intensive technologies such as AI algorithms will have **enormous energy requirements**. This demand should be reduced as far as possible through the research and development of energy-efficient technologies (Arunov & Scholz, 2023) and covered by renewable sources (Yan Zhang et al., 2023). In order to avoid wasting energy, however, the utility value of individual applications in the metaverse must always be considered: the focus should **not be on technology push,** **but on user pull** when developing and implementing technologies in order to address the real needs of users and leverage new economic potential for urban society.

At the same time, it should be mentioned at this point that the urban metaverse has enormous potential for **saving resources and avoiding emissions**, for example by avoiding duplication of work (Hölzle et al., 2023) and physical travel (Arunov & Scholz, 2023), virtualizing non-essential products such as toys (Allam et al, 2022) and processes such as product development (Hölzle et al., 2023), as well as by increasing infrastructure resilience (Raes, Ruston McAleer & Kogut, 2022) and the quality of life and health of citizens (Guckenbiehl et al., 2021). All of these value drivers can become part of the overall value proposition of dedicated business models for the urban metaverse. We explore these in more detail in the following chapter for the broad area of planning, constructing and operating urban buildings and infrastructure.

# 4. Deep Dive: Use Cases in the Lifecycle of Public Buildings and Infrastructures

**Many German cities are facing numerous planning and construction projects**. In addition to catching up on missed maintenance investments in roads and drainage/utility structures to maintain the status quo, a technology region like **Aachen** is currently still dealing with the expansion of stationary and mobile broadband (fiber optics, 5G, etc.). Germany's westernmost city is also part of the **Rhenish mining region**. In the wake of the political decision to phase out coal, lignite mining regions are facing enormous transformation challenges. This structural change from coal mining and power generation to alternative energy sources also requires huge infrastructure investments (Bundes-regierung, 2023). This structural change is embedded in the larger political paradigm of energy and mobility transformation. To reduce greenhouse gas emissions, decentralized renewable energies are to be expanded and electrification is to be promoted. In the transportation sector, for example, space is needed to charge the growing number of electric cars, but also to provide bike and car-sharing stations for new forms of private transportation. This means making existing traffic areas more bicycle-friendly and replacing parking garages with green spaces. At the same time, cities need to adapt to climate change, for example by implementing "sponge city" concepts to better deal with heavy rainfall events.

Almost all areas of a city's infrastructure are therefore facing transformation challenges, and there will be many changes in supply and disposal infrastructures such as electricity, district heating, gas and water. All these transformations represent great opportunities for the economy and society (and are essential for a future worth living in the long term), but for the citizens of a city they initially mean construction sites, torn-up roads, detours, traffic jams and frustration.

In this context, a major **value proposition of the urban metaverse is to rethink the planning and implementation of these infrastructure measures**. In this chapter, we outline a vision of how digital twins, XR technologies and AI can be used beneficially for public works in cities in the future under the umbrella of the metaverse.

| Table 2: Value Propositions of the Urban Metaverse Along the Lifecycle of Urban Facilities and Infrastructures | | |
|---|---|---|
| **Phase of the lifecycle** | *View of city management and (external) specialist staff (architects, project managers, trades, etc.)* | *View of the affected stakeholders, in particular citizens and companies* |
| **Planning** | **Increasing administrative efficiency**<br>▪ Increased efficiency in administration and improved decision-making through smoother communication and coordination between departments and external stakeholders<br>▪ Improvement of long-term maintenance planning through continuous data availability and updating as well as integration of data silos<br><br>**Increase in technical planning quality**<br>▪ Context-dependent simulation and optimization, e.g. in the area of 5G infrastructure<br>▪ Simulation of the influence of new buildings in existing buildings (e.g. densification), thereby increasing the efficiency of land management<br>▪ Better energy planning based on digital twins; possibility of energy optimization as early as the planning phase<br><br>**Increase in planning acceptance**<br>▪ Possibilities for immersive visualization and experience-oriented communication of planned construction measures<br>▪ Opportunities for qualified user feedback and inclusion of user needs in planning<br>▪ Scalable, efficient option to include an integrated user journey from the perspective of specific stakeholder groups in the planning process | **Increasing participation and involvement**<br>▪ Simpler and more inclusive opportunities for participation in urban planning processes<br><br>▪ Increasing the quality of use of the buildings or infrastructures to be planned, as specific needs can be better communicated and, if necessary, incorporated<br><br>**Increasing the efficiency of stakeholders affected by construction measures**<br>▪ Better ability to assess the consequences of a measure and align own processes accordingly (e.g. outdoor catering areas at a construction site; route planning for delivery services)<br><br>▪ Better opportunities for long-term investments, as e.g. the future situation and residential quality of a street can be visualized and communicated in advance |
| **Construction and Execution** | **Reducing costs and avoiding additional costs**<br>▪ Increased efficiency through better coordination of various construction measures and avoidance of duplication of work thanks to a standardized information basis for all parties involved<br>▪ Avoidance of complexity costs by simulating the influence of planned | **Increasing participation and involvement**<br>▪ Improving proximity to citizens and transparency through interactive and participatory platforms<br><br>▪ Increased acceptance through participation in construction progress, but |
| | changes in the construction process and the possibility of coordinating these with all parties involved on the basis of a uniform data model<br>■ Reducing costs and time by avoiding accidents and damage to existing infrastructure, e.g. by making existing pipelines visible to construction personnel using AR glasses<br><br>**Improving the working conditions and efficiency of those carrying out the work**<br>■ Shortening training times and increasing the quality of execution by visualizing work steps<br>■ Avoidance of errors and accidents and thus acceleration of the implementation process through visual and coordinated instructions in AR<br>■ Increasing the quality of execution and better coordination by integrating technical experts in the event of problems or archaeological finds, for example | also better understanding of the complexity and challenges of construction measures<br>■ Possibility for citizens to exercise control and screening rights regarding the efficient use of taxpayers' money<br><br>**Increasing the efficiency of stakeholders affected by construction measures**<br>■ Better planning of your own processes thanks to the availability of up-to-date information, e.g. on traffic routing or access to parking spaces during a construction project<br><br>**Increasing the acceptance of construction measures**<br>■ Options for immersive visualization and experience-oriented information on construction progress and ad-hoc changes |
| **Operation and Use** | **Increasing the availability and preventing outages of urban infrastructures**<br>■ Increased efficiency through continuous monitoring and control of buildings or units in real time<br>■ Increase in service quality by minimizing downtimes through predictive maintenance based on digital twins of units<br><br>**Reduction in the operating costs of existing infrastructures**<br>■ Enabling economies of scale and bundling effects through data aggregation and thus optimizing the operation of infrastructures<br>■ Increasing the capacity or utilization of an existing infrastructure by simulating traffic flows, for example, and better defining regional mobility concepts | **Improving the quality of life and stay**<br>■ Improved quality of life through better (coordinated) planning in regular operations, fewer unplanned outages and faster reactions in the event of disruptions to urban infrastructure<br>■ Increased participation and better planning of own processes through the provision of urban real-time information, e.g. traffic situation, parking spaces, pollen count, utilization of leisure facilities and sports facilities<br>■ Increasing participation and quality of urban infrastructure through citizen sensor projects |

| Phase of the lifecycle | *View of city management and (external) specialist staff (architects, project managers, trades, etc.)* | *View of the affected stakeholders, in particular citizens and companies* |
|---|---|---|
| | **Increasing decision-making competence**<br>▪ Increasing decision-making competence by moving from a reactive attitude (Where are the problems? How do I fix them?) to a proactive attitude (How can problems not arise in the first place?)<br>▪ Obtaining critical empirical planning information for subsequent periods through continuous target/actual comparisons and better insights into actual (and not just assumed) usage patterns<br>▪ Signaling innovation leadership and competence as a region through efficient digital administrative processes and stakeholder centricity; promoting business relocations<br><br>**Increased resilience and reaction speed in the event of disruptions or disasters**<br>▪ Increasing the quality of planning and implementation by simulating incident scenarios and virtual exercises for dealing with these scenarios<br>▪ More efficient and scalable training and education of emergency personnel and decision-making bodies in city management for disaster situations<br>▪ Better planning of defusing, protection and evacuation measures by simulating a bomb disaster during the defusing of aerial bombs from the Second World War | **Increasing personal resilience and reducing the impact of disruptions or disasters**<br>▪ Possibility of training citizens and other stakeholders on how to behave in the event of massive disruptions, (natural) disasters or terrorist incidents<br><br>▪ Possibility of immersive digital communication and interaction in the metaverse in the event of non-availability of physical infrastructure (e.g. schools or cultural institutions in the event of a pandemic, etc.) |

**Table 2: Value drivers of the urban metaverse along the lifecycle of urban buildings and infrastructure**

To this end, in this chapter we present a series of use cases based on the lifecycle of public buildings. We look at **three levels**: the individual building, the individual infrastructure and the city as a whole, each within the **lifecycle phases** of planning (design), construction, and use & operations, including monitoring and maintenance. [9] Some of the applications presented below are in development or even hypothetical. Others, however, already exist.

We take **two perspectives** on the discussion of the use cases: **A) the value proposition of the urban metaverse from the point of view of the management and (contracted, external) professional staff of a city** planning and executing construction measures (architects, project managers, construction workers, etc.), and **B) its value proposition from the point of view of the affected stakeholders**, in particular citizens and firms.

**Table 2 provides a summary overview of the different value drivers** that together can define the value proposition of the urban metaverse along the lifecycle of urban buildings and intra-structures. A similar argument can be made for other domains of the metaverse, such as education, tourism, shopping, culture, political communication and citizen participation in local politics.

## Planning & Design Phase

Working with virtual prototypes and simulation systems in the first lifecycle phase, the planning of public buildings and infrastructure, is not new per se. The main value proposition of the urban metaverse lies in the improved availability of more holistic and up-to-date planning data (especially the context and environment of the project) as well as other possibilities for cooperation and communication with various stakeholders.

A first value driver of the urban metaverse from the **perspective of (planning) specialists**, i.e. for civil engineers and architects, is a larger creative solution space at the **building level**. For example, different materials and daring building styles can be creatively tried out without restrictions and, above all, tested in their future environment. The resulting model designs can be tested without risk prior to implementation (Hölzle et al., 2023) and explored via VR. In this way, resources can be saved and errors avoided, especially in complex construction projects, if future users can test them already in the planning phase (Vessali et al., 2022).

An energy performance certificate and a building resource passport can also be issued for buildings as early as in the planning phase, as envisaged by the German government (Baur, 2024). This will allow the $CO_2$ footprint to be measured over a building's lifetime. The City of **Helsinki**, for example, uses its digital twin "Energy and Climate Atlas" to track its climate targets. It measured the energy characteristics of individual buildings, which are primarily determined by the materials used and the energy sources used. For buildings owned by the city, heat and electricity consumption is even tracked life and continuously, creating a real **virtual energy twin**. This is particularly valuable for Finnish cities due to the cold-temperate climate. In addition, the energy input can be modeled for each hour of the day via solar radiation and shading (similar projects for solar potential analysis exist in the cities of **Hannover** and **Bremen**). By also comparing heating alternatives, recommendations for energy optimization can be made. For example, it is conceivable that several houses could join forces and share the costs of a borehole for the use of geothermal energy.[10] The US city of Ithaca in the state of New York is pursuing a similar approach (Socio, 2022).

---

[9] The life cycle phases of dismantling, recycling and disposal are excluded.

Planning in virtual space is also valuable for the design of **infrastructure**. For example, the Swedish telecommunications company Ericsson used the metaverse platform NVIDIA Omniverse to investigate the interaction between 5G antennas and urban objects such as buildings and trees in the digital twin of a city. This enabled the network to be optimally designed in terms of performance and network coverage (Kerris, 2021). Underground infrastructures for supply and disposal can also be virtually simulated and planned (Vessali et al., 2022).

At **city level**, digital twins can also be used for three-dimensional holistic planning and testing - and here both new cities and urban districts on the greenfield as well as existing cities (Kuru, 2023). This is particularly valuable for larger cities that are densifying areas and thus growing in height and depth rather than width. For example, it can be used to illustrate how a new high-rise building would affect the skyline.

For barrier-free urban development, for example, obstacles for people with disabilities can also be identified at an early stage by AI agents with the same behavior moving around the virtual city and recognizing problems (Guckenbiehl et al., 2021). The simple adaptation of such a "user journey" to the needs of different stakeholders - and their simultaneous interaction with it from different perspectives - holds great potential for better urban development.

Major cities such as **Amsterdam**, **Singapore**, **Zurich** and **Stockholm** are already using their digital twin to efficiently plan land management (Bentley Systems, 2021; Cattaneo, 2023). Bremen models the urban microclimate, taking into account buildings and greenery, which depicts heat development, thermals and air currents in order to take urban heat islands into account in further urban development. The dispersion of pollutant particles through the air currents can also be simulated.[11] In addition, lighting options can be simulated (**Hannover**), the noise emissions of new wind turbines can be analyzed and the visual axes of special buildings can be planned (**Bremen**).

---

[10]  https://www.hel.fi/en/decision-making/information-on-helsinki/maps-and-geospatial-data/helsinki-3d

[11]  https://vc.systems/mikroklimasimulation-in-bremen//

**Stakeholders and users** of the infrastructure to be planned, such as citizens, politicians and investors, can also use the urban metaverse to inform themselves and participate in urban development. We take up this idea below and describe the value drivers based on the **participation levels** of information, consultation, involvement and co-determination.

In a first step, stakeholders can obtain **information** about currently planned projects and construction measures. The digital models of urban planning can be presented publicly in early project phases and can be explored by stakeholders via various end devices in 3D environments (see below) - the next evolutionary stage of 2D online maps for citizen information, so to speak. The visualization and contextualization of planned projects is particularly accessible for viewers, especially in complex processes such as construction projects, as it is easier to recognize connections. It is also easier to understand why the construction measures are necessary and sensible. Overall, this allows for more effective communication (Guckenbiehl et al., 2021). For example, it can illustrate what a new building will look like and how it will be embedded in its surroundings (Kusuma & Supangkat, 2022). In this way, the best locations for new stores, cafés and schools can be identified in advance, for example by modeling the frequency of pedestrian traffic.

In addition to pure information, the digital twin can also serve as a dialog platform for stakeholder **consultation and participation** in digital planning processes. Change requests can be communicated and taken into account at an early stage, resulting in less rework. Misunderstandings can be avoided and the overall acceptance of construction projects can be increased. The Swedish cities of **Gothenburg** and **Stockholm** communicate planned changes in urban areas via a dialog platform linked to the digital model, which can be accessed online via private devices such as smartphones or PCs. In Gothenburg, for example, a tunnel construction project was visualized in this way (Bentley Systems, 2021).

The highest levels of participation can be achieved by empowering stakeholders to **actively participate**. Critical design decisions can be made jointly in the urban meta-verse, which leads to greater acceptance and interest compared to classic urban construction projects. Finally, all these opportunities for interaction also offer a field of experimentation for the co-design of a city. Citizen competitions are conceivable, for example, in which residents can directly modify planning drafts or collaboratively develop innovative design drafts (Allam et al., 2022). In this way, residents can actively contribute to urban design and, if necessary, even vote democratically on the submitted designs.

These high levels of participation are made possible by AR and VR interfaces. Using **virtual reality (VR)**, planning designs and concepts can also be better interpreted by laypeople - i.e. anyone who finds it difficult to draw conclusions about a complex three-dimensional reality on the basis of two-dimensional construction drawings. Users can playfully explore buildings or infrastructure in the planning and construction phase without any safety instructions or helmets and discuss them directly with urban planners in virtual space. This could get younger people in particular interested in seemingly boring topics such as infrastructure projects in the utilities and waste disposal industry and develop an interest in relevant professions. As an alternative to VR, **augmented reality (AR)** can make it possible to experience planning designs in a concrete context

"on site" using regular smartphones or tablets. In **Cologne's** Ehrenstraße, citizens have already been able to use an AR app from the company Cityscaper [12] on their devices to get a first impression of the future image of the street, which was projected onto a large construction site on the street. The data was provided by the planning architecture firm and shows the result of a car-free, green recreational area in 3D. The actual redesign was also decided with the citizens (Baumanns, 2023). Urban planning designs are also visualized in the City of **Bremen's** AR app "BremenGo" (Gellhaus, 2024). [13]

For broader social participation, citizens can also be invited to face-to-face events to discuss the respective construction project at a **digital planning table**. In the Digital Participation System (DIPAS) project, an interactive touch table shows the city in both two and three dimensions and tells a story (in text form) depending on the perspective. In addition, citizens can give direct feedback on planning procedures and are informed about other participation formats. The software of the same name also ensures a uniform experience across all devices and generates statistics and key figures.

For those who use and are affected by an urban construction project, the various applications therefore enable **higher levels of participation**, from information, consultation and involvement to co-determination and co-creation. Again, this is not an achievement of the urban metaverse, but has long been common (and legally required) practice in many municipalities. However, the metaverse makes this participation scalable, inclusive, and experiential. As a result, the acceptance and understanding of necessary construction measures should increase. Most importantly, the quality of planning could be improved if the knowledge and needs of many user groups could be more easily incorporated into the planning process, a typical pattern of open innovation and co-creation with customers. This is especially true for shops and restaurants -- commercial stakeholders that are often particularly affected by construction. Restaurants, for example, can use a metaverse application to better assess how a planned construction project will affect their outdoor space - or perhaps make new outdoor spaces possible in the future.

## Construction and Execution Phase

Urban metaverse users can also benefit during the construction implementation of the planning designs. In particular, **construction coordination** with different **specialist personnel** and companies can be made smoother and more resource-efficient. At **building level**, the work steps can be defined with integrated project planning software and distributed to the units carrying out the work (civil engineering, bricklayers, electricians, painters, etc.) with fixed dates. In the event of plan changes, all those involved are informed directly via the digital twin - but always have coordinated access to the same planning status and changes made, especially during execution. The effects of changes to the original planning can also be discussed directly with all trades - based on a precise simulation of the effects of the planning.

---

[12] The cityscaper company is also taking part in the KomIT project.
[13] https://www.bremen.de/bremen-go

It should also be emphasized here that the most important prerequisite for realizing this vision is not the urban metaverse, but rather networked digital twins of planning and implementation that are available to all parties involved in real time and in every detail. This is precisely the vision of Building Information Modeling (BIM), a trend that has dominated the construction industry for years, even if the level of implementation still varies greatly. However, BIM will become an important part of the urban metaverse when combined with its other technological design factors.

This is also transferable to the **infrastructure level**, where project managers can also check whether the planned construction measure clashes with the schedules of other companies. Ideally, companies can implement construction measures together in this way and avoid duplicating work and resources. At **city level**, for example when building entire cities such as the Saudi Arabian city "**The Line**", construction coordination is becoming increasingly complex. An urban metaverse can help to maintain an overview and distribute work effectively. By incorporating data that could be relevant to the construction project (traffic flow, weather, etc.), the quality of project management can be further increased.

In addition, construction site personnel can carry out **training and remote training** with XR equipment before and during the actual construction phase. This helps to test dangerous and complex work steps such as the transportation and installation of heavy loads (e.g. wind turbine rotor blades) without risk. In addition, construction workers can be briefed on the respective work step in augmented or virtual reality at the start of their work.

It is also conceivable that construction workers could wear lightweight AR glasses that visualize the next work step. In this way, (as yet) unskilled workers such as trainees and temporary **workers on the construction site** could be instructed. Although temporary work on construction sites is prohibited in Germany, this could prevent accidents in countries with less stringent regulations. Pictorial and easy-to-understand instructions on the job can help people learn new processes faster, improve communication, and avoid errors and accidents. This makes the implementation process faster and safer. At the **building level**, for example, an electrician can use lightweight glasses to view the electrical installations planned and stored in the digital model directly on site and install them according to plan. During renovations, the electrician can see which cables go where.

This can also be applied to construction measures at **infrastructure level**. Here, the glasses can show the 3D section of the subsoil, i.e. the cables and pipes that were either mapped directly in the digital model of the subsoil when they were laid or, if necessary, were recorded three-dimensionally in civil engineering during a previous construction project. This means that an excavator driver can now "look underground" and dig the hole more quickly without causing costly damage. Solutions for 3D recording of underground structures are already offered by companies such as Pix4D. [14] Finally, construction work can also be tracked live by city management at **city level** from a distance using AR glasses or in virtual space using VR glasses.

During the work, **specialists and decision-makers** can also be **consulted** for various problems and issues. In the case of unknown finds at **building and infrastructure level**, for example, archaeologists can be called in quickly and easily on site without having to travel to the site. The specialists can have live access to the image capture of the AR glasses or the 3D laser scan of the site. Any archaeological finds such as walls, vessels and jewelry can be virtually documented, archived and forwarded to virtual museums, for example. If a large number of historical structures and finds in the city have been virtualized, it may even be possible to recreate the city historically. This is an exciting possibility, especially for cities with a long history. In this way, contemporary history can be made sustainably accessible and preserved without endangering the real artifacts through (mass) tourism (Allam et al., 2022). At **city level**, the rapid connection of administrative staff to current construction measures ensures greater efficiency.

In the meantime, **affected stakeholders** can use their AR and VR devices (or the digital planning table) to find out about the construction measures. At **building and infrastructure level**, for example, it is important for stakeholders to know when which construction steps will take place and how long they will take. Until now, citizens have usually (if at all) received an information letter with a drawing that provides information about the expected time frame and the type of planned changes - for example, a renewal of the water and wastewater network and house connections. In the future, an additional QR code could be placed in the letter or directly at the construction site, allowing citizens to look underground using an AR-enabled device (for example, via a balcony or directly at the construction site fence).

The clear visual representation makes it easier for stakeholders to understand the necessity of the construction measure. In addition, those concerned can better adjust to a temporary everyday life with the construction site, for example by means of daily updated information on traffic routing and accessibility to parking spaces, which may change during the construction work. At **city level**, AR glasses wearers can use the live information to better navigate around the construction work in the city.

## Operations and Utilization Phase

Once construction is complete and operations have commenced, **specialist personnel** can regularly inspect, monitor, control and maintain the public structures. At **building and infrastructure level**, structures that are difficult to access, such as bridges or television towers, can be inspected on a recurring basis using sensor-equipped drones.

---

[14] https://www.pix4d.com/de/produkt/rtk/

During the **inspection**, damage can be noted in the respective digital twins in the urban metaverse and a repair can be ordered. Time series data from IoT devices, which can be visualized virtually, is particularly suitable for **continuous monitoring and control in real time**. For example, a janitor can use lightweight AR glasses to see at a glance whether all rooms in the university have been automatically locked and the lights switched off in the evening. A sewer network operator can call up operating data such as the flow rate in the pipes and the water levels in the buffer tanks, but can also be visually alerted in the event of faults such as failed pumps. Finally, they can intervene in the control process themselves via the glasses' gesture recognition. In **Shanghai**, the underground infrastructure of an entire district, the Liang district, was already digitally mapped during construction and is continuously monitored (Cai et al., 2023).

At **city level**, existing digital models of buildings and infrastructure can be combined into a larger digital model of the city, which is fed with a wealth of dynamic real-time data from IoT devices and upgraded to a networked digital twin. In this way, it will be possible to react quickly and perhaps even automatically to live conditions in the city. The data can come from a variety of urban areas and stakeholders, including private companies, public institutions such as authorities and universities, and citizens (White, Zink, Codecá, Clarke, 2021). Among other things, this enables the joint monitoring and control of several buildings and infrastructures in the city, such as water supply and wastewater disposal.

If the (real-time) data is recorded continuously, changes and transformation processes can be shown **retrospectively**. For example, while different usage patterns can be analyzed over time in terms of water consumption at **building level**, the data on domestic and commercial water consumption can be further processed at **infrastructure level**, for example to determine how much wastewater is produced in the catchment area each year. Taking other data into account, such as the weather, it can be shown that there are more peak loads in the wastewater network due to heavy rainfall. At **city level**, the effectiveness of individual measures can then be checked with concrete figures in the respective digital twin, for example whether shower bans can help in the event of acute water shortages or whether sewer network controls lead to better utilization and less environmental pollution (to be checked here in the digital twin of the receiving water body).

**Forecasts** can also be created using data-driven simulations of future events **at all levels of the city**. For example, maintenance and renovation requirements can be estimated in advance. This concept of predictive maintenance originates from the industrial sector. We will look at the data-driven simulation of fictitious events in the next subsection.

Working with historical, current and forecast data makes it possible to set up **command centers** and achieve operational excellence with the help of AI-supported recommendations for action (Vessali et al., 2022). In this way, the transition from a reactive attitude (Where are the problems? How do I fix them?) to a proactive attitude (What can be done to prevent problems from arising in the first place?) can be achieved. A metaverse-enabled city could also measure KPIs such as urban $CO_2$ emissions, but also the quality of life and health of citizens based on anonymized health data (Guckenbiehl et al., 2021).

It is very challenging to set up such comprehensive data-driven analyses and forecasts and to manage the city across departments. To do this, it must be possible to link the data from different areas of the city and understand and model the interdependencies of urban processes. The vision of an urban metaverse can only be realized through a broad database and interdisciplinary and cross-sectoral cooperation.

**Citizens and other stakeholders** can also enter data into the system as a kind of human sensor technology (citizen sensors). For example, citizens can actively participate in the removal of street damage, graffiti, garbage and fallen trees by reporting to the city administration quickly and smoothly via corresponding platforms (Crowley, Curry & Breslin, 2013). There are already portals such as FixMyStreet[15] or the **Aachener Mängelmelder**[16], which offer this service. In the INFRASense project of the city of **Osnabrück**, even the bicycles of some citizens are temporarily equipped with sensors that record the surface condition of the paths and at the same time collect data on the mobility behavior of citizens such as speed and waiting times. The collected data is transferred to the BIQUEmonitor platform, where it is analyzed and displayed. Citizens without sensors can provide feedback on road quality directly on the platform.[17] Such ideas can be further developed in the urban metaverse. Citizens can pass on information directly to the digital twin via augmented or virtual reality or purchase their own IoT-enabled objects (e.g. bicycles) that are connected to the digital twin. Some cities, such as **Dublin**, have conducted initial tests in this area (White et al., 2021).

Conversely, citizens can also benefit from real-time data. Many cities already provide their citizens with a lot of information online. For example, the current capacity utilization of roads and parking lots, as well as air and water quality, can be displayed. In future, citizens will be able to see this data in real time on their devices, for example on their car's head-up display when searching for a parking space.

## Simulation of What-if Scenarios

The simulation of fictitious events affects all three lifecycle phases of the planning, implementation and use of public buildings - a use case that is primarily located in the field of **disaster prevention and safety**. Cities can encounter a "Black Swan" at any time - a previously unexpected, improbable event with major consequences. The terrorist attacks of September 11, 2001, the **Fukushima** nuclear disaster on March 11, 2011 or the severe weather events and flooding in the Ahr-valley are often cited as examples of a Black Swan. The emergence of digital twins gives cities the opportunity to prepare for such uncertainties and thus increase their resilience.

---

[15] https://www.fixmystreet.com/
[16] https://maengelmelder.aachen.de/
[17] https://www.infrasense.de/

As part of various vulnerability analyses, what-if scenarios can be simulated realistically and dangers can be uncovered at an early stage. Emergencies such as terrorism, but also natural disasters such as earthquakes, floods, fires and even tsunamis can be simulated risk-free in various forms and intensities. It is also possible to simulate heat waves and sandstorms (Vessali et al., 2022). The prerequisite for such simulations is that all relevant areas of the city can be modeled interactively.

This enables urban planning to record and evaluate safety requirements and react on the basis of the new information (Allam et al., 2022). If, for example, the critical areas where water masses collect in the event of heavy rainfall are known, targeted construction measures can be taken to protect these areas. Assembly areas can be lowered and at the same time serve as retention basins for the water floods. Sealed surfaces can be replaced by green areas in the sense of a sponge city concept, which can absorb a lot of water in a short time and release it again with a time delay. Conversely, the effectiveness of existing climate adaptation plans can also be validated (Allam et al., 2022).

In addition, emergency personnel and decision-making bodies such as city management can prepare for disasters. The international criminal police organization Interpol is already simulating police operations for training purposes in a virtual environment (Interpol, 2024). Even if an incident actually occurs, a city remains capable of acting via the command center. The civilian population can be warned in good time and evacuated with visual support, for example by projecting the optimal escape route onto the ground. (Kuru, 2023). The company Virtual City Systems, for example, offers the simulation of a bomb detonation, which depicts the propagation of the blast wave and the fragmentation flight. This is particularly valuable when defusing aerial bombs from the Second World War in order to identify danger zones and plan defusing, protection and evacuation measures.[18]

In addition to these safety-related aspects, there are **numerous other possible use cases for simulations** and corresponding planning approaches -- not only in the context of one city, but especially in the interaction of several cities (in the sense of the EU vision of a Cityverse presented above). For example, it is not only possible to simulate the effects of different construction options on inner-city traffic flows, but also to examine, test and improve inter-city, i.e. regional and national mobility concepts in simulations. As the example of the networked digital twin of Deutsche Bahn in Chapter 3 shows, there are important parallels between the industrial and urban metaverse. Simulations can improve the responsiveness and resilience of urban infrastructure, which in turn improves the quality of life for citizens (Kusuma & Supangkat, 2022).

---

[18] https://vc.systems/loesungen/urbane-simulation/

# 5. Conclusion: The Urban Metaverse Has Just Started to Emerge

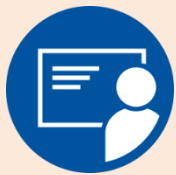

*"The metaverse in the smart city is nearer than we think. [But] there is no chance for cities to make an impact if big platforms like Uber and Airbnb dictate their business model. The same can happen with the metaverse! So cities and municipalities have to act now!"*

Harmen van Sprang, Co-founder & CEO, Sharing Cities Alliance & Studio Sentience

We already live in a world where digital resources are often as valuable as physical resources -- from the money in your bank account or the digital images on your smartphone to NFT-backed digital art. We see the metaverse as a continuation of this trend. It is an extension of the physical world by one, or rather many, virtual layers and a continuation of physical activities in the digital world, with a constant feedback loop between the worlds.

The pioneers of urban metaverse initiatives present impressive visions and ambitious plans. Cities such as **Seoul, Dubai, Tampere, Singapore, Amsterdam** and **Helsinki** are pioneers who are using the potential of the metaverse to improve the quality of life of their citizens -- but also to improve their image in the quest for **innovation leadership in the global competition among metropolitan areas**.

**The benefits of these projects are undeniable**. Virtual platforms allow for broader and more inclusive citizen participation. VR and AR technologies also offer new opportunities for immersive learning environments and improve the skills of the population. Digital twins and simulations optimize planning processes, help to avoid errors and reduce costs, as we have seen in our deep dive into the planning, implementation and operation of urban buildings and infrastructure.

Real-time data and simulations support cities in implementing environmental protection measures and adapting to climate change. By creating metaverse hubs and fostering innovation ecosystems, cities aim to unlock new economic growth opportunities and create virtual jobs. Our table of urban metaverse value drivers at the beginning of Chapter 4 and the discussion in this chapter provide many points of reference for potential value propositions that investments in an urban metaverse can realize for all stakeholders.

**Nevertheless, there are many challenges** that need to be overcome. First and foremost, the acceptance of citizens and urban users is crucial to the success of any metaverse initiative. Many people are still unfamiliar with the new technologies and need support to be able to use them effectively. Far too many applications are **purely technology-centric projects** for the sake of doing a showcase, but do not really address the benefits for citizens and other stakeholders. With the "jobs-to-be-done" approach and Outcome Driven Innovation, we have described above an approach for aligning future urban metaverse applications more closely with value drivers for all stakeholders. Not everything that is currently being positioned as a future technology under the term the industrial or urban metaverse is new. Digital twins, urban data spaces, BIM, simulation techniques and VR technology have been around for a long time and can already provide great benefits in isolation -- yet their degree of adaptation is sometimes very low. Perhaps what is needed to overcome the existing barriers to diffusion is the bundling effect and the holistic applications that this makes possible.

Many of the initiatives presented are still in the planning or early implementation stages, and **concrete results and measurable successes are lacking**. Interoperability between different systems and platforms remains a major challenge, and the technical foundations for seamless integration are often not yet fully developed. There is also a lack of standardization and binding norms. Important issues of accessibility and participation remain unresolved. Security and the privacy are also key concerns that have not yet been fully addressed.

From an operator perspective, the **development of sustainable business models** for the urban metaverse is still in its infancy. Many projects are currently heavily dependent on public funding and it is unclear how the ongoing operating costs will be covered once the initial investment has been used up. Despite the wide range of use cases and considerable investment, it is therefore clear that there are still many hurdles to overcome before the visions and value propositions of the urban metaverse can be fully realized.

However, it is clear that the **development of the metaverse can have a powerful impact on the development of cities**. If many work and leisure activities do indeed shift to virtual space and there are therefore fewer incentives to live in the city, migration could return to rural areas in the long term, triggering an urban exodus (Allam et al., 2022; Kuru, 2023). "Why should I spend my time in traffic jams and queues, or fly to the other side of the world for a whole day, when I can experience all this at home if I just put on my glasses? Why should I live in a city with expensive rents and little space?", many city dwellers will ask themselves. Entertainment venues such as cinemas, theaters and amusement arcades as well as office complexes could disappear, and travel infrastructure such as roads and airports could be cut back. However, the savings in space and energy will be offset by the high energy consumption of the metaverse and its technologies, which will place new demands on our energy systems (Allam et al., 2022).

It is crucial that cities continue to invest in research and development, develop clear strategies and business models and actively involve citizens in the process of the smart city and its urban metaverse. This is the only way to realize the full potential of the metaverse to create sustainable and innovative cities of the future. However, they should not let the sceptre be taken out of their hands and wait for American or Chinese corporations to promise the "urban metaverse as a service". Just as the privatization of public infrastructure requires **a balance between private investment from operators and public investment**, the urban metaverse requires a balance between the use of global platforms, operating systems and wearables and local (or national) initiatives.

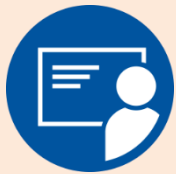

*"No matter which scenario, it is evident that to leverage the potential of the metaverse fully, organizations must exhibit considerable creativity and avoid merely replicating real-world elements in the virtual space. This approach entails optimizing the unique aspects of both realities. It mirrors the early days of the Internet, where companies often failed by not adapting their processes and business models to the new digital environment. The metaverse requires a similar innovative approach. …*

*There is consensus that the potential of the Metaverse and the risk of missing out on its opportunities are too significant to ignore completely. Building sufficient expertise within an organization today will be advantageous for navigating tomorrow's landscape if these environments gain prominence. Better to be safe than sorry."*

Andreas Kaplan, KLU Hamburg, und Michael Haenlein, ESCP Business School, Paris (Source: Kaplan & Haenlein 2024)

However, cities and municipalities in particular should be able to meet these challenges better than other actors. In a recent article, Clough & Wu (2024) use the example of the historical development of the Strip in **Las Vegas** to show many parallels to the development of a future metaverse. In their analysis, the two authors draw many analogies as to how it might be possible to create a consensus-based urban metaverse in a way that is both cooperative and inclusive, as well as profitable.

After all, the key challenges that managers and companies face in implementing the metaverse are the same ones that urban planners face in developing new cities and neighborhoods: Long-termism, openness to technology, participation, focus on the common good, interaction and cooperation are not new concepts for urban planners (as they are for many companies), but are part of everyday life. We therefore appeal to all city and community leaders to be open to the opportunities and potential of the urban metaverse and to focus on their existing strengths and competencies when implementing it, so that the vision of the urban metaverse can become a living reality.

# About Us

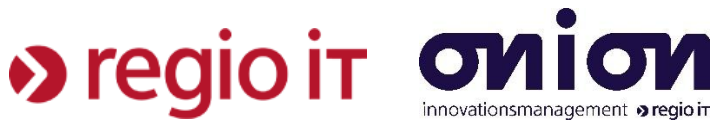

As the largest municipal IT service provider in North Rhine-Westphalia, regio iT GmbH is the ideal IT partner for public clients: for municipalities and schools, energy suppliers and waste disposal companies as well as non-profit organizations. Headquartered in Aachen and with branches in Gütersloh and Siegburg, regio iT offers strategic and project-related IT consulting, integration, IT infrastructure and full service in four business areas: IT Service & Operations, Administration & Finance, Energy & Waste Management, Education & Development. It is involved in numerous research and development projects as well as nationwide initiatives for new technologies. With around 720 employees, it currently supports more than 56,000 customers and 300 schools and school administrations.

ONION INNOVATION: REGIO IT'S INNOVATION MANAGEMENT is forward-looking and focused on developing innovative solutions to the challenges of digitalization. Together with all those who like to do "0816", we explore how innovation works beyond marketing. Our hypothesis: Together, across the boundaries of organizations, hierarchies, disciplines and cultures, with the courage to experiment, directly, openly, appreciatively and in constant dialogue, we can achieve more than incremental progress - maybe even leaps. We believe in the innovative power of networks. They are a source of ideas and knowledge - whether local, sectoral or technology-based. Together, ONION is changing the way we see things - with companies, cities, local authorities, committees, people who think and think ahead, with the digital community.
https://www.regioit.de

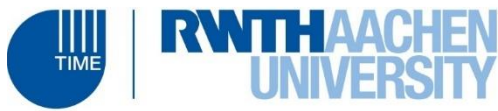

At the Institute for Technology and Innovation Management (TIM) at RWTH Aachen University, an interdisciplinary team of seven professors and postdocs, about 30 research assistants, many student researchers and a large network of dedicated visiting professors and guest researchers investigate current issues in strategic technology and innovation management. The research and project work of RWTH TIM focuses on the transformation of established organizations in the face of disruptive technological innovations such as Industry 4.0, Artificial Intelligence, Smart Products & Services or Additive Manufacturing; the systematic development and evaluation of new business models; Open Innovation and the development of successful regional innovation ecosystems to promote sustainability and economic development; as well as the development and improvement of instruments and metrics for effective technology management and technology policy.
https://www.tim.rwth-aachen.de

# Urban Metaverse: The Smart City in the Industrial Metaverse

## Potential of the metaverse for real-time capable, interactive and inclusive infrastructure applications in the city

The urban metaverse describes an immersive 3D environment that connects the physical world of the city and the citizens living in it with its digital data and systems. Physical and digital reality merge and open up new possibilities for the design and use of the city. This study serves as an impulse generator and guide for decision-makers in cities and municipalities, urban planners, IT experts, companies and anyone interested in the future of urban spaces in order to understand the opportunities and challenges of the urban metaverse as a further development of the smart city and to set the course for sustainable and innovative urban development. To this end, the study analyzes the opportunities that the urban metaverse offers for urban management and the everyday lives of citizens, presents key technologies, and highlights the socio-economic challenges of implementation. The focus is on the potential of the urban metaverse to optimize the planning and operation of urban infrastructures, to promote inclusion and citizen participation, and to enhance the innovative capacity of cities and communities. The study makes four recommendations for the implementation of metaverse applications in the urban context: 1. Focus on user-centered design, 2. Ensure ubiquitous accessibility, 3. Proactively shape the regulatory framework, and 4. Develop viable business models.

**A trend study by the Institute for Technology and Innovation Management at RWTH Aachen University on behalf of regio iT gmbh**